\documentclass[twocolumn,superscriptaddress,amssymb,amsmath,nobibnotes,aps,
prd,showpacs,nofootinbib]{revtex4}%
\pdfoutput=1

\usepackage{}%
\usepackage{graphicx}
\usepackage{epsf}
\usepackage{bm}
\usepackage{amsmath}
\usepackage{amsfonts}
\usepackage{amssymb}
\usepackage{epstopdf}%
\usepackage{color}

\begin{document}

\title{Observational constraints on sign-changeable interaction models and alleviation of the $H_0$ tension}

\author{Supriya Pan}
\email{supriya.maths@presiuniv.ac.in}
\affiliation{Department of Mathematics, Presidency University, 86/1 College Street, 
Kolkata 700073, 
India}

\author{Weiqiang Yang}
\email{d11102004@163.com}
\affiliation{Department of Physics, Liaoning Normal University, Dalian, 116029, P. R. 
China}

\author{Chiranjeeb Singha}
\email{cs12ip026@iiserkol.ac.in}
\affiliation{Department of Physical Sciences, 
Indian Institute of Science Education and Research Kolkata,
Mohanpur 741246, West Bengal, India}

 \author{Emmanuel N. Saridakis}
\email{msaridak@phys.uoa.gr}
\affiliation{Department of Physics, National Technical University of Athens, Zografou
Campus GR 157 73, Athens, Greece}
\affiliation{Department of Astronomy, School of Physical Sciences, University of Science 
and Technology of China, Hefei 230026, P.R. China}


\begin{abstract}
We investigate various scenarios which include interaction forms between dark matter and 
dark energy that exhibit sign reverse, namely where the 
transfer of energy between the dark fluids changes sign during   evolution. We study 
the large-scale inhomogeneities in such interacting scenarios and we confront them with  
the latest astronomical data. 
Our analysis shows that the sign-changeable interaction models are able to produce stable 
perturbations. Additionally, the  data seem to slightly favor a non-zero 
interaction, however, within $1\sigma$ confidence level (CL) 
the scenarios cannot be distinguished from non-interacting cosmologies. We find 
that the best-fit value of the dark-energy equation-of-state parameter lies in the 
phantom 
regime, while the quintessence region  is also allowed nevertheless at more than  
2$\sigma$ CL. Examining the effect of the interaction on the CMB TT and matter 
power spectra we show that while from the simple spectra  it is hard to 
distinguish the interacting case from $\Lambda$CDM  scenario, in the residual graphs  
 the interaction is indeed traceable. Moreover, we find that sign-changeable interaction 
models can reconcile the $H_0$ tension, however the $\sigma_8$ tension is still 
persisting. Finally,  we examine the validity of the laws of thermodynamics and we 
show that  the generalized second law is 
always satisfied, while the second derivative of the total entropy becomes 
negative at late times which implies that the universe tends towards thermodynamic equilibrium.
\end{abstract}


\maketitle
\section{Introduction}

Recent observations of various origin suggest that around 69\% of the universe consists 
of the dark energy sector, while around  26\% constitutes of the cold dark matter one  
\cite{Ade:2015xua}. One usually assumes that these two sectors do not mutually
interact, and the resulting scenario is capable of describing very efficiently the 
various 
sets of independent observations. Nevertheless, since the underlying microscopic theory 
of both these sectors is unknown, there is not any field-theoretical argument against the 
consideration of a possible mutual interaction. Furthermore, such an interaction could 
alleviate the known coincidence problem \cite{Amendola-ide1, 
Pavon:2005yx, delCampo:2008sr}, namely why are the energy 
densities of  dark  matter and dark energy  currently of the same order although they 
follow completely different scaling laws during evolution. Hence, in the literature one
can find many such interacting scenarios (see \cite{DE_DM_1,Wang:2016lxa} for review and 
references therein), independently of the specific dark-energy nature, namely whether
it arises from fields   \cite{Copeland:2006wr,Cai:2009zp} or
through a gravitational modification 
\cite{Nojiri:2010wj,Capozziello:2011et,Nojiri:2017ncd}.
The interacting scenarios can be very efficient in describing late-time universe, and 
moreover they seem to be slightly favored comparing to non-interacting ones
\cite{Basilakos:2008ae,Basilakos:2009ch,Salvatelli:2014zta,Basilakos:2012ra,Sola:2016jky,
Xu:2013jma,
Costa:2013sva,Basilakos:2013vya,Ruiz:2014hma,Cai:2015zoa,Li:2015vla,Valiviita:2015dfa, 
Eingorn:2015rma,Pan:2012ki,Nunes:2016dlj,Caprini:2016qxs,Yang:2017ccc, Yang:2017zjs}. 
  Additionally, they seem to be efficient in addressing the $H_0$ tension 
\cite{Kumar:2017dnp,DiValentino:2017iww,Yang:2018euj,Yang:2018uae,Kumar:2019wfs} as well 
as the 
$\sigma_8$ tension 
\cite{Kumar:2019wfs,vandeBruck:2017idm,Thomas}.

In the above scenarios the interaction term, that determines  the interaction rate 
and thus the flow of energy between the dark matter and dark energy sectors, is 
introduced phenomenologically. Although there is a large variety in such choices, 
they are usually assumed to have the same sign, namely the energy flow is from one 
component to the other during the whole universe evolution. However, an interesting 
question arises, namely what would happen if  we allow for a sign change of the 
interaction term during the cosmological evolution. 
 Such a consideration might be further useful to investigate the 
dark sectors' physics.
Hence, in the present work we desire to investigate this possibility and in particular to 
confront the obtained scenarios with different observational data coming from probes like 
the cosmic microwave background radiation (CMB), supernovae type Ia (SNIa), baryon 
acoustic 
oscillations (BAO), and Hubble parameter measurements.

We organize the present work in the following way. In Section \ref{sec-2} we provide the 
cosmological equations at background and perturbative levels, in presence
of arbitrary coupling in the dark sector. In Section \ref{sec-models} 
we present the models that we wish to study in this work. In Section \ref{sec-data} 
we describe the observational datasets and the fitting methodology. In Section 
\ref{sec-results} we 
provide the constraints on the models, and we perform a Bayesian analysis   
 in comparison with $\Lambda$CDM cosmology.  Finally, we close the present work in 
Section \ref{summary} with a brief summary of the results.

\section{Interacting cosmology}
\label{sec-2}

In this Section we briefly review interacting cosmology. We consider a homogeneous 
and isotropic flat Friedmann-Robertson-Walker (FRW)
line element of the form
\begin{eqnarray}
ds^2 = -dt^2+a^2(t)\,
\delta_{ij} dx^i dx^j,
\end{eqnarray}%
where $a (t)$ is the expansion scale factor of the universe.
Furthermore, we consider   the universe to be filled 
with baryons, cold dark matter, radiation, and the dark energy sector (which may be of 
effective origin or not), all of which considered as barotropic perfect fluids.
 Thus, the Friedmann equations that determine the 
 universe evolution    are written as
\begin{eqnarray}
&& H^2   = \frac{8\pi G}{3} \rho_{t},\label{f1}\\
&& 2\dot{H} + 3 H^2  = - 8 \pi G\, p_{t}\label{f2},
\end{eqnarray}
 with  $G$ the Newton's constant and $H=\dot{a}/a$ the Hubble 
function (dots denote derivatives with respect to $t$). In the 
above equations we have 
introduced the total energy density and pressure respectively  as
$\rho_{t} = \rho_r +\rho_b +\rho_c 
+\rho_x$ and $p_{t} = p_r + 
p_b + p_c + p_x$, with the subscripts $r,\; b,\; c,\; x$ denoting  
radiation, baryon, cold dark matter and   dark energy.
 
Although the Bianchi identities lead to the conservations of the total energy momentum 
tensor, they do not imply anything for the separate sectors, and thus one can assume that 
some of them mutually interact \cite{Amendola-ide1, 
Pavon:2005yx, delCampo:2008sr,Faraoni:2014vra}. In this work we allow the dark 
matter and dark energy sectors to interact, while radiations and baryonic matter are 
considered to be conserved. In particular, we assume that
\begin{eqnarray}
&&\dot{\rho}_b + 3 H \rho_b =0\\
&&\dot{\rho}_r + 4 H \rho_r =0\\
&&\dot{\rho}_c + 3 H  \rho_c = -Q,\label{cons-cdm}\\
&&\dot{\rho}_x + 3 H (1+w_x)\rho_x = Q,\label{cons-de}
\end{eqnarray}%
where $w_x$ is the dark-energy equation-of-state parameter (for baryonic and dark matter 
we consider the dust case $w_b=w_c= 0$, while for radiation as usual $w_r=1/3$).
The introduced quantity $Q$ is a phenomenological descriptor of the 
interaction, and its form is considered arbitrarily. If $Q>0$ then the energy transfer is 
from cold dark matter (pressureless dark matter) to dark energy, while if $Q<0$ then it 
is from dark energy to dark matter. Moreover, as usual the conservation equations for 
baryonic matter and radiation give $\rho
_{b}=\rho _{b0}a^{-3}$ and   $\rho _{r}=\rho 
_{r0}a^{-
4}$  respectively, with $\rho _{i0}$ ($i=r,b$)   the value of $\rho
_{i}$ at present time. 
 In summary,  if the interaction function is given  then the Friedmann equation 
(\ref{f1}) 
alongside the conservation  equations 
(\ref{cons-cdm}) and (\ref{cons-de}), can determine the evolution of the
universe.

One can see that the conservation equations (\ref{cons-cdm}) and (\ref{cons-de}) can be 
written in 
an alternative way as 
\begin{eqnarray}
&&\dot{\rho} _{c}+3H\left( 1+w_{c}^{\mathtt{eff}}\right) \rho
_{c} =0, \label{cons-cdm1} \\
&&\dot{\rho} _{x}+3 H\left( 1+w_{x}^{\mathtt{eff}}\right) \rho
_{x} =0, \label{cons-de1}
\end{eqnarray}
where   $w_{c}^{\mathtt{eff}}$, and 
$w_{x}^{\mathtt{eff}}$ 
are the effective equation-of-state parameters for cold dark matter and dark energy, 
given as
\begin{eqnarray}
&&w_{c}^{\mathtt{eff}} =\frac{Q }{3 H\rho _{c}}, \\
&&w_{x}^{\mathtt{eff}} =w_{x}-\frac{Q }{3 H\rho _{x}}. 
\end{eqnarray}%
Hence, as we observe, the interaction  affects the  equation of state 
of these components. In particular, dark matter may depart from dust while dark 
energy may be quintessence or phantom like even if the initial $w_x$ is fixed to one 
regime.

We proceed to the investigation of the above scenarios at the level of perturbations.
We consider    scalar perturbations around an FRW metric given by 
\cite{Mukhanov,Ma:1995ey,Malik:2008im}
\begin{eqnarray} 
\label{perturbed-metric} 
&&
\!\!\!\!\!\!\!\!\!\!\!\!\!\!\!\!\!\!\!
ds^{2} =a^{2}(\tau )\left\{-(1+2\phi )d\tau ^{2}+2\partial _{i}Bd\tau 
dx^{i}\right.\nonumber\\
&&
\ \ \ \ \ \ \ \,
\left.
+\left[(1-2\psi )\delta _{ij}+2\partial _{i}\partial _{j}E\right]
dx^{i}dx^{j}\right\},
\end{eqnarray}%
where $\tau$ represents the conformal time and the quantities $\phi$, $B$, %
$\psi$, $E$, denote the gauge-dependent scalar perturbations. Thus, in the case of 
interacting cosmology with $w_x \neq -1$ the perturbation equations in 
the 
synchronous gauge ($\phi
=B=0$, $\psi =\eta $, and $k^{2}E=-h/2-3\eta $), with $k$ the Fourier mode and $h, \eta$, 
being the 
metric scalar perturbations \cite{Ma:1995ey},   
are written as  
\cite{Valiviita:2008iv, Majerotto:2009np,Clemson:2011an}:
 \begin{eqnarray}
&&
\!\!\!\!\!\!\!\!\!\!\!\!\!
\delta _{x}^{\prime } =-(1+w_{x})\left( \theta _{x}+\frac{h^{\prime }}{2}
\right) -3\mathcal{H}w_{x}^{\prime }\frac{%
\theta _{x}}{k^{2}}     \notag \\
&& -3\mathcal{H}(c_{sx}^{2}-w_{x})\left[ \delta _{x}+3\mathcal{H}%
(1+w_{x})\frac{\theta _{x}}{k^{2}}\right] 
\notag \\
&&+\frac{aQ }{\rho _{x}}\left[ -\delta _{x}+\frac{\delta Q  }{Q  }+3\mathcal{H}%
(c_{sx}^{2}-w_{x})\frac{\theta _{x}}{k^{2}}\right] ,
\label{perteq1}
\\
&&
\!\!\!\!\!\!\!\!\!\!\!\!\!
\theta _{x}^{\prime } =-\mathcal{H}(1-3c_{sx}^{2})\theta 
_{x}+\frac{%
c_{sx}^{2}}{(1+w_{x})}k^{2}\delta _{x}
\notag \\
&&
+\frac{aQ }{\rho _{x}}\left[ \frac{%
\theta _{c}-(1+c_{sx}^{2})\theta _{x}}{1+w_{x}}\right] , \label{theta-x}
\\
&&
\!\!\!\!\!\!\!\!\!\!\!\!\!
\delta _{c}^{\prime } =-\left( \theta _{c}+\frac{h^{\prime 
}}{2}\right) +%
\frac{aQ  }{\rho _{c}}\left( \delta _{c}-\frac{\delta Q  }{Q  }\right) , 
\label{eqn:delta-c}\\
&&
\!\!\!\!\!\!\!\!\!\!\!\!\!
\theta _{c}^{\prime } =-\mathcal{H}\theta _{c}.
\label{perteq4}
\end{eqnarray}
In the above expressions we have introduced the overdensities  $\delta _{i}=\delta \rho 
_{i}/\rho _{i }$,  as well as the velocity perturbations $\theta_i$, with 
    primes denoting   derivatives with respect to the conformal time $\tau$, with 
$\mathcal{H}=\frac{a'}{a}$   the conformal Hubble function. Moreover, $c_{sx}^2$ is the 
adiabatic sound speed, which in the following will be set to 1
\cite{Valiviita:2008iv,Majerotto:2009np, Clemson:2011an} (the adiabatic sound speed for 
the dark matter in the dust case is $c_{sc}^2 =0$).  

 Now, for  the case where $w_x=-1$, namely in the case of interacting 
vacuum 
scenario, 
the perturbation equations are slightly different. In the synchronous gauge the momentum 
conservation equation for dark matter reduces to $\dot{\theta}_c =0$ \cite{Wang:2014xca}, 
while the density perturbations can be recast into \cite{Wang:2014xca} 
\begin{eqnarray}
\dot{\delta}_c = -\frac{\dot{h}}{2} + \frac{Q}{\rho_c} \delta_c~.
\end{eqnarray}
However, in this gauge  the vacuum energy is spatially 
homogeneous, i.e. $\delta \rho_x = 0$.  

 Finally, it is important to remark that although in the present work, as 
well as 
in many works in the literature, the spatial curvature of the universe is not 
considered,  a series of works   shows that the observational data 
do not exclude the non-zero 
curvature  
\cite{Ryan:2019uor,Mitra:2019rzc,Park:2018tgj,Park:2018fxx,Ryan:2018aif,
Park:2018bwy,Yu:2017iju,Farooq:2016zwm,Chen:2016eyp}, and hence the general universe 
picture should include the spatial curvature as a free parameter. We will address such a 
complete investigation in a separate work.

\section{Sign-changeable Interaction Models}
\label{sec-models}

Since the appearance of interacting cosmology,  a variety of phenomenological coupling 
functions $Q$ have been introduced and  investigated  in the literature. In general, most 
of the coupling functions indicate a particular direction of energy transfer, namely
either the transfer of energy from dark matter  to dark energy or vice versa. This 
includes   functions of the form $Q \propto H \rho_c$, $Q
\propto H \rho_
x$, $Q  \propto H (\rho_c + \rho_x)$, $Q  \propto H (\rho_c \rho_x)/(\rho_c + 
\rho_x)$,  $Q  \propto H\rho_x^2/(\rho_c + \rho_x)$, nonlinear forms, etc
\cite{Billyard:2000bh,Barrow:2006hia,Zimdahl:2006yq,
Saridakis:2007wx,He:2008tn,Quartin:2008px,Chen:2008ft,
Chimento:2009hj,Jamil:2009eb,Valiviita:2009nu, Gavela:2010tm,Baldi:2011th,
Thorsrud:2012mu,Pan:2013rha,Yang:2014hea,Yang:2014gza,Nunes:2014qoa,Pan:2014afa,
Chen:2011cy,Pan:2012ki,Li:2013bya,Kofinas:2016fcp,
Mukherjee:2016shl, Pan:2016ngu, Yang:2017yme,Sharov:2017iue, Pan:2017ent, 
Yang:2018ubt,Yang:2018pej,
 Yang:2018euj,vonMarttens:2018iav,Yang:2018qec,Asghari:2019qld,Ikeda:2019ckp}.

However, in principle one could also have the case in which the interaction 
function, i.e. $Q$, changes sign during the evolution, namely the energy transfer between 
the interacting sectors changes direction \cite{Sun:2010vz, Wei:2010cs, 
Forte:2013fua,Guo:2017deu,Arevalo:2019axj}. 
In the following subsections we examine two of such models separately.

\subsection{Interaction function  $Q = 3 H \xi (\rho_c - \rho_x)$}

The first interacting model in which the interacting function changes sign during 
evolution is 
\begin{eqnarray}\label{IDE1}
Q (t) = 3 H \xi (\rho_c - \rho_x), 
\end{eqnarray}
where $\xi$ is the coupling parameter. Note the interesting feature that  when $\rho_c = 
\rho_x$ then the interaction becomes zero even if $\xi \neq 0$. This kind of feature was 
observed 
in a different interaction model 
\cite{Yang:2018xlt}. 
\begin{figure}[ht]
\includegraphics[width=0.4\textwidth]{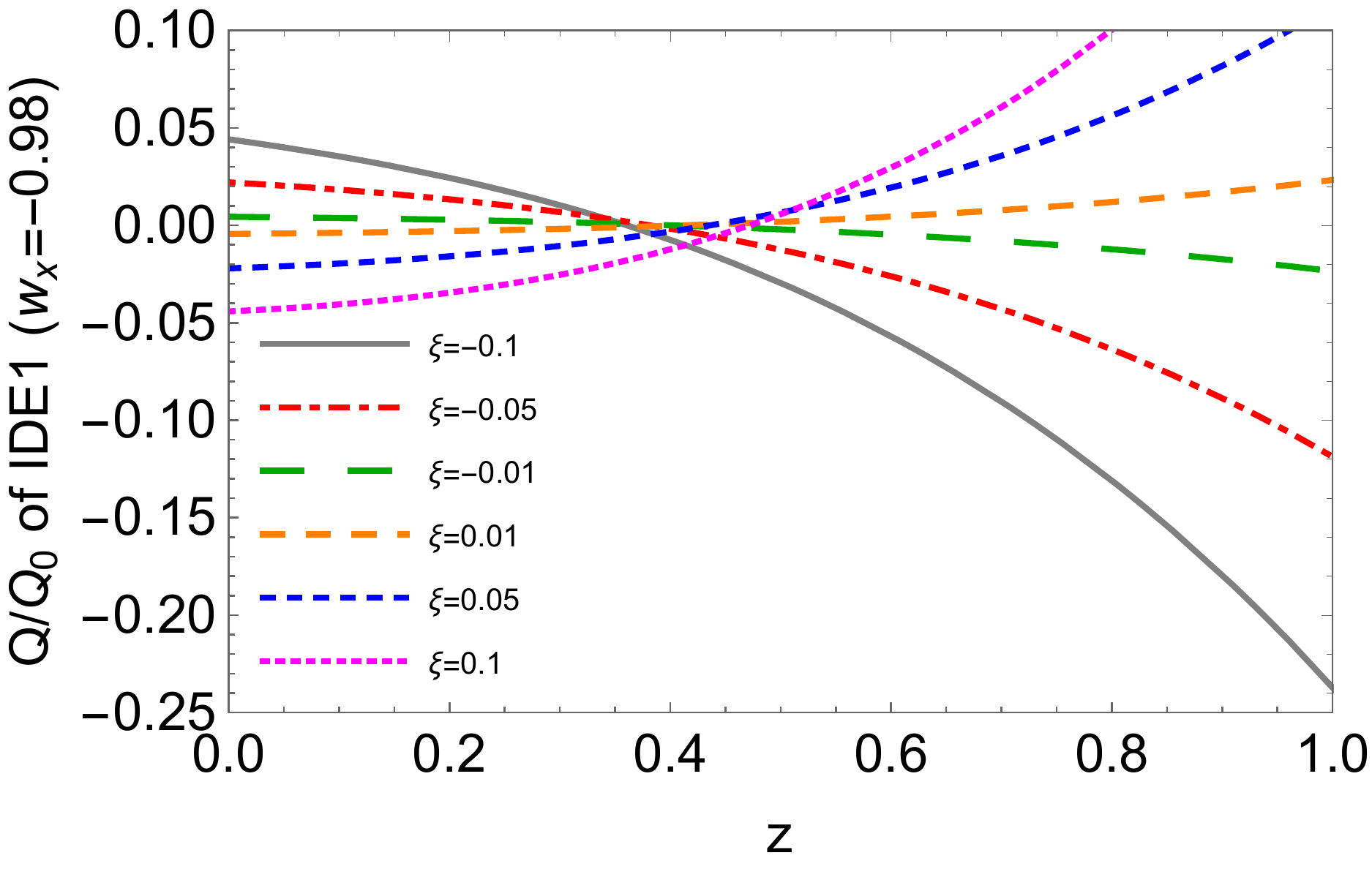}
\includegraphics[width=0.4\textwidth]{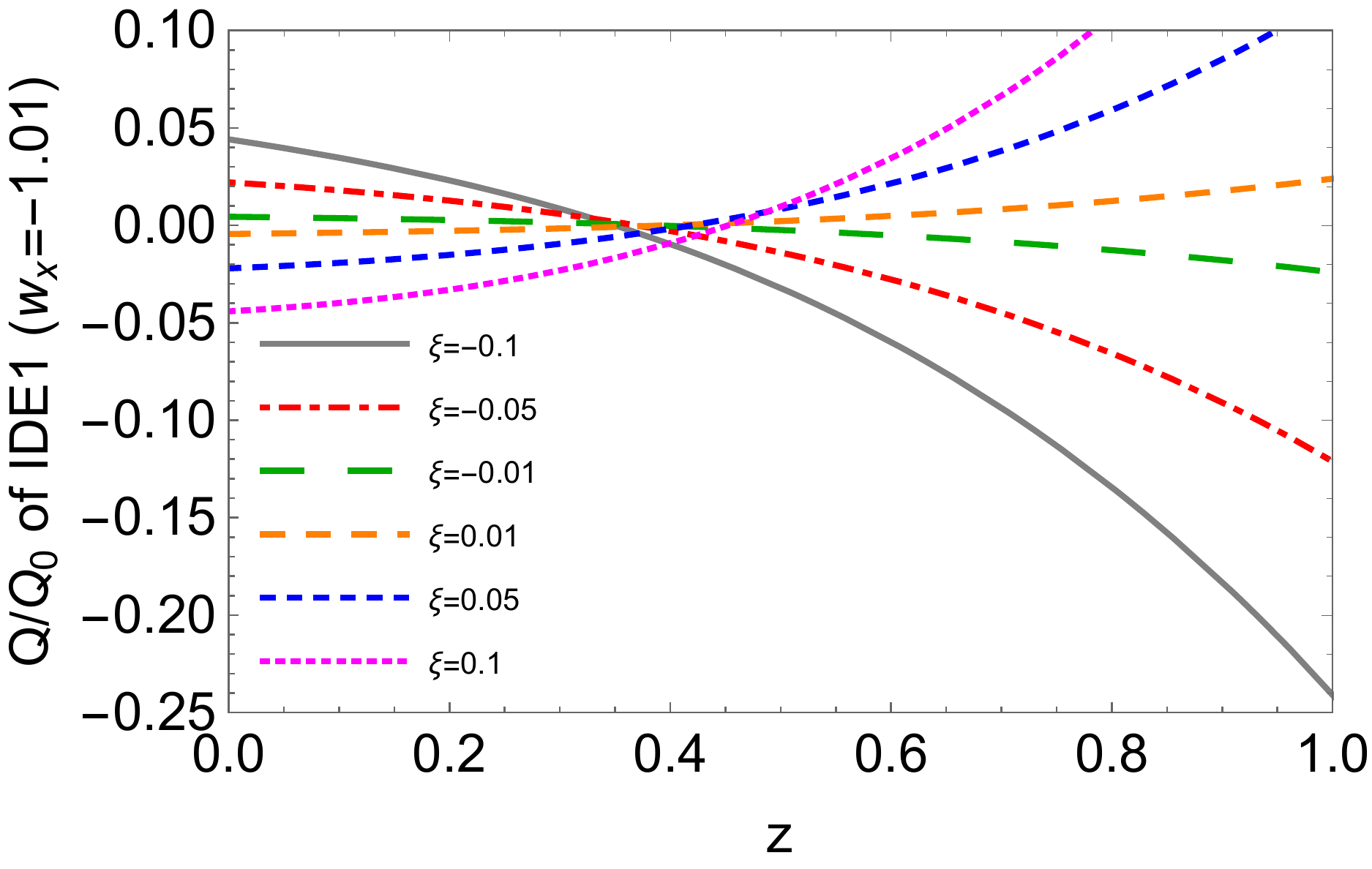}
\caption{{\it{The evolution of the normalized interaction function $Q/Q_0$, where
$Q_0 = H_0 
\rho_{t0}$ (with $H_0$ and $\rho_{t0}$ the present values of the Hubble parameter and the 
total energy density 
$\rho_t$ respectively),
for the interaction model IDE1 of (\ref{IDE1}), for 
various values of the coupling parameter $\xi$ in units where $8\pi G=1$,  for $w_x  
=-0.98$ (upper graph) and for  $w_x = 
-1.01$ (lower graph).}}}
\label{fig:Q-IDE1}
\end{figure}
\begin{figure}[ht]
\includegraphics[width=0.4\textwidth]{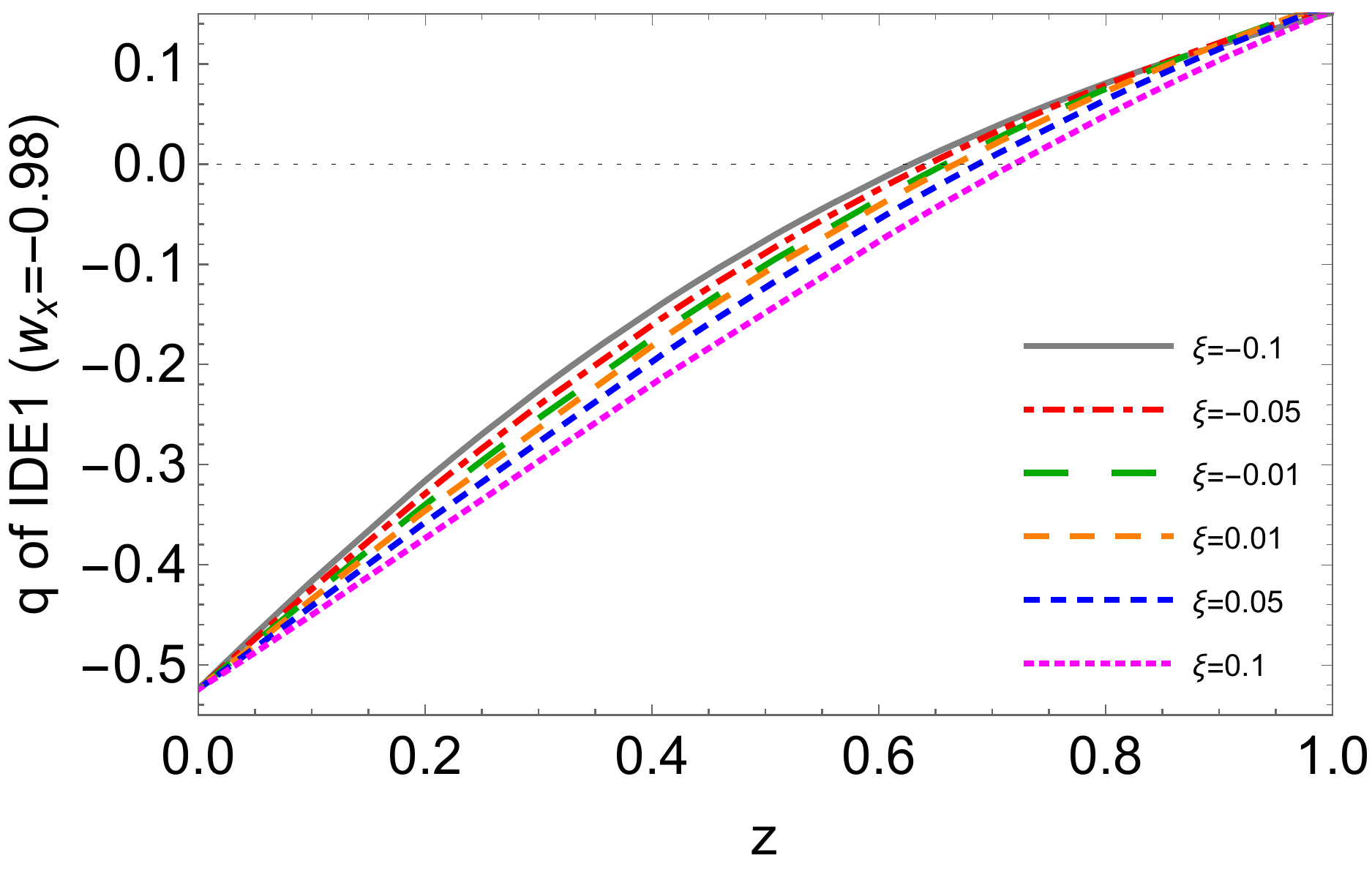}
\includegraphics[width=0.4\textwidth]{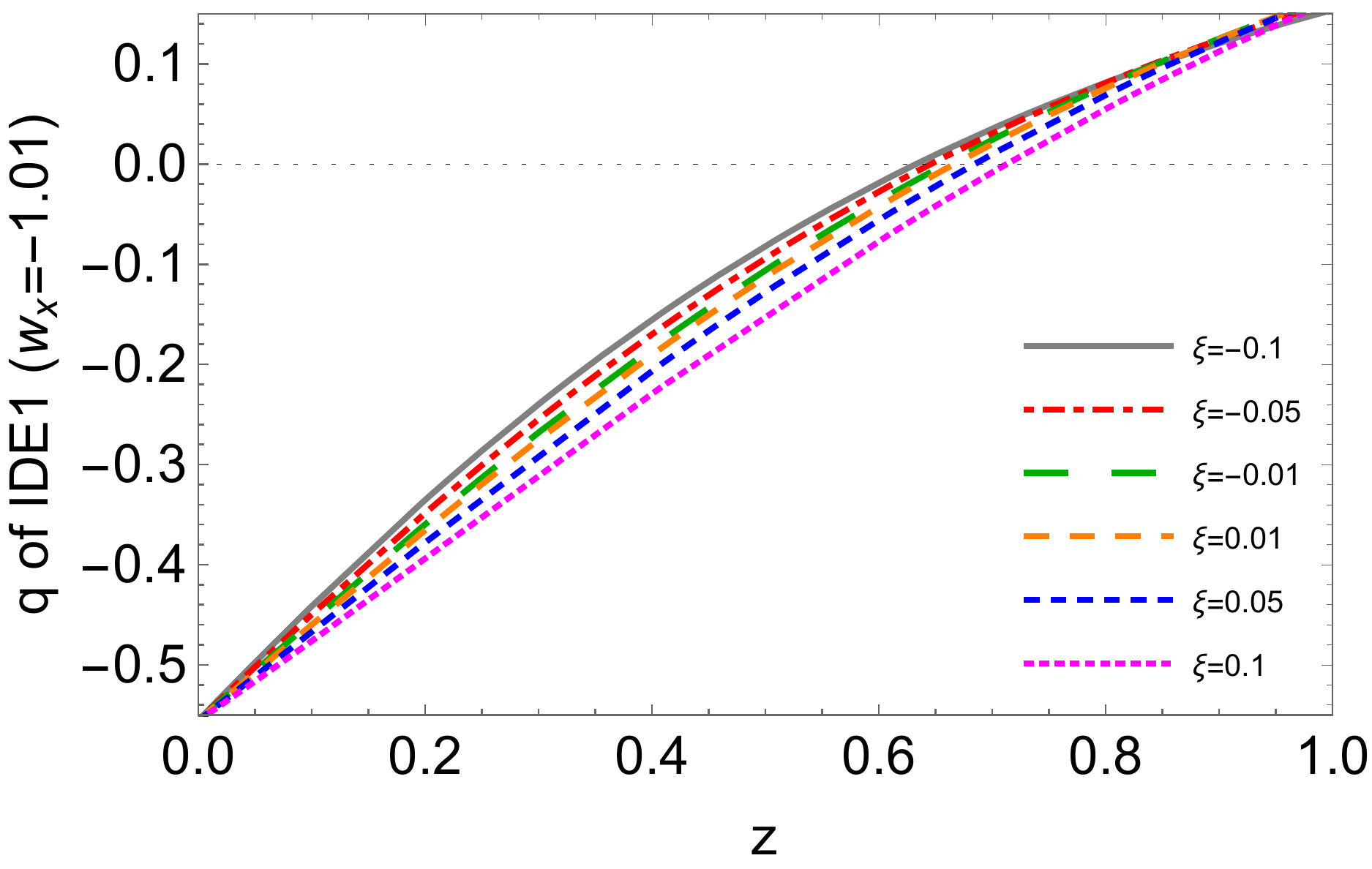}
\caption{{\it{The evolution of the deceleration parameter as a function of the redshift, 
for the interaction model IDE1 of (\ref{IDE1}), for $w_x  =-0.98$ (upper graph) and for  
$w_x = 
-1.01$ (lower graph), for 
various values of the coupling parameter $\xi$ in units where $8\pi G=1$. We have set
 $\Omega_{x0}\approx0.69$, $\Omega_{c0}\approx0.25$, 
$\Omega_{b0}\approx0.05$ and $\Omega_{r0}\approx10^{-4}$ in agreement with observations.
  }}}
\label{fig:q-IDE1}
\end{figure}
Here, we consider two choices for 
the dark energy equation-of-state parameter $w_x$, namely the cosmological constant 
case $w_x= -1$ (which corresponds to the interacting vacuum scenario, from now on 
model IVS1), and the case $w_
x \neq -1$ (from now on model IDE1).  The latter case is interesting since it allows 
to extract constraints on the dark energy equation-of-state parameter and examine whether 
it lies in the quintessence    or in the 
phantom regime.

We proceed by numerically elaborating 
the 
background cosmological equations, using as independent variable the redshift, $z = 
a_0/a-1$, 
setting the present scale factor $a_0$ to $1$. Moreover, we introduce the density 
parameters of the various components through $\Omega_i=8\pi G \rho_i/(3H^2)$ and we 
set their current values as $\Omega_{x0}\approx0.69$, $\Omega_{c0}\approx0.25$, 
$\Omega_{b0}\approx0.05$ and $\Omega_{r0}\approx10^{-4}$ in agreement with observations 
\cite{Ade:2015xua}. 

In order to obtain a picture for the above  interaction form,
in Fig. \ref{fig:Q-IDE1} we first depict the evolution of $Q$ normalized by $Q_0 = H_0 
\rho_{t0}$ (with $H_0$ and $\rho_{t0}$ the present values of the Hubble parameter and the 
total energy density 
$\rho_t$ respectively), for various values of $\xi$. As we can see, the sign-change 
direction and its exact moment depends on $\xi$, while  
the qualitative features of the interaction function do not change for  $w_x> -1$ 
or 
$w_x < -1$.

\begin{figure}
\includegraphics[width=0.39\textwidth]{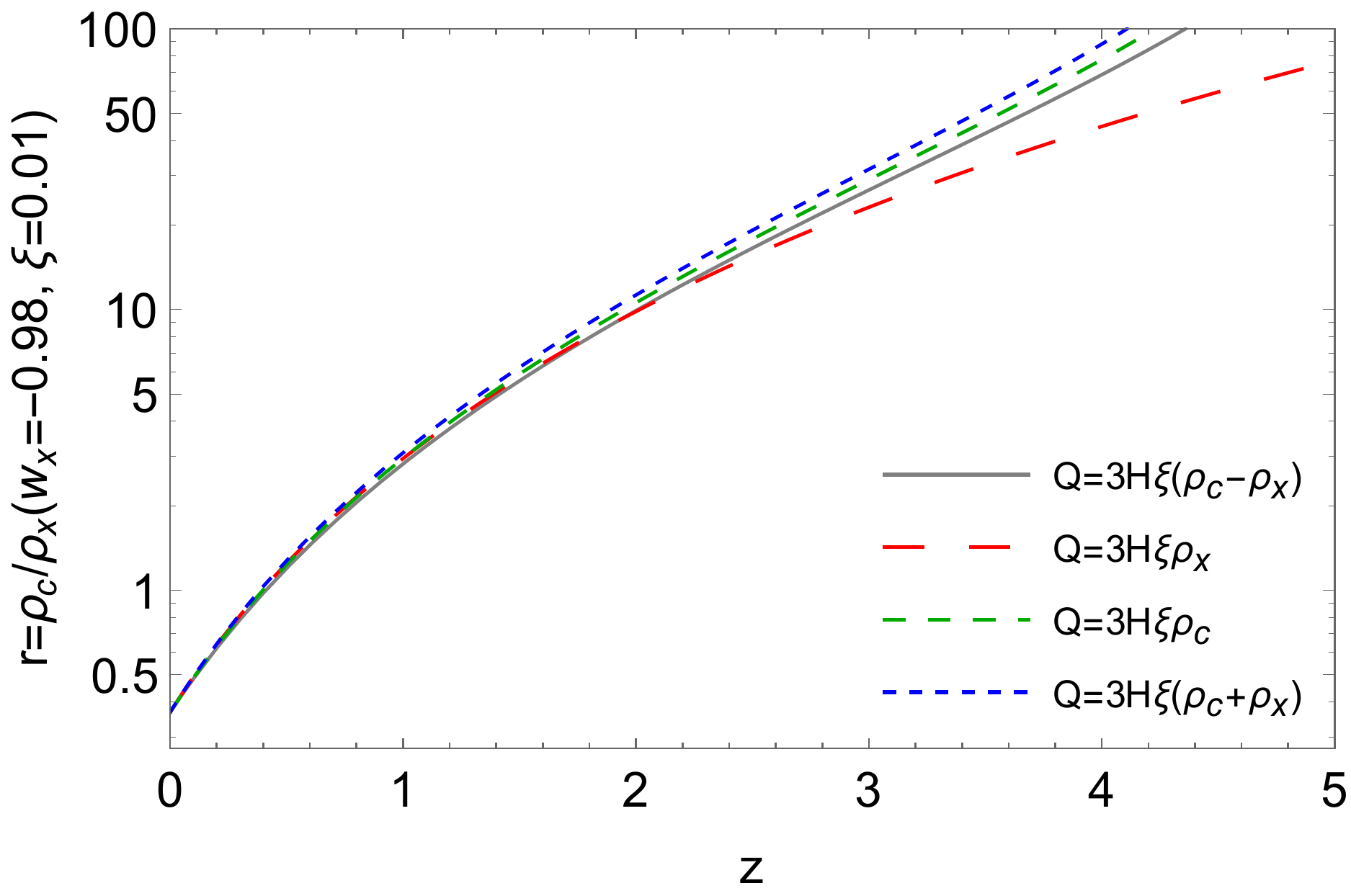}
\includegraphics[width=0.39\textwidth]{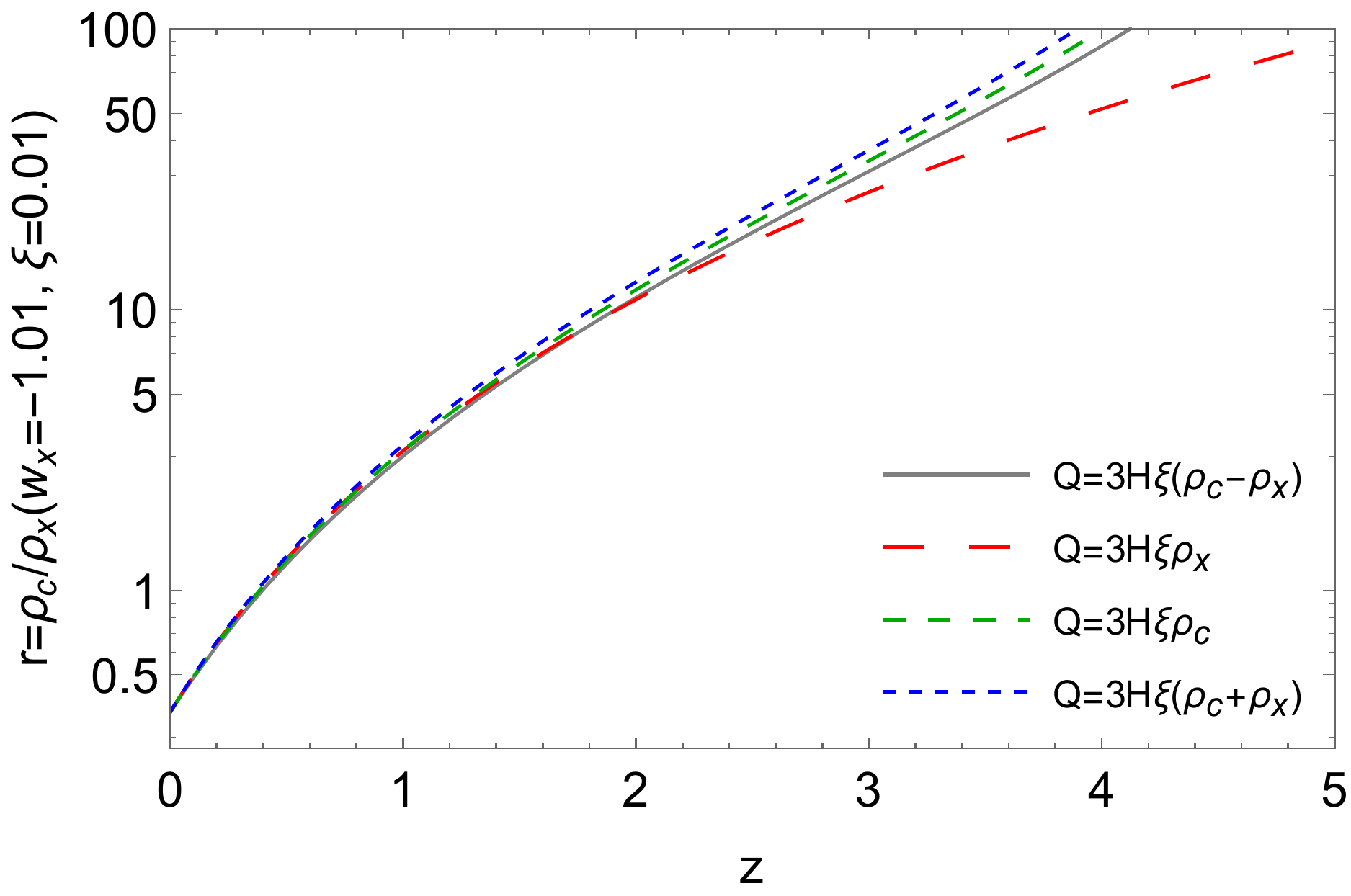}
\caption{\it{
The evolution of the coincidence parameter $r = \rho_c/\rho_x$,
for the interaction model IDE1 of (\ref{IDE1}),   for $w_x  
=-0.98$ (upper graph) and for  $w_x = 
-1.01$ (lower graph), setting  a 
typical 
value of $\xi = 0.01$ in units where $8\pi 
G=1$.
For comparison we additionally depict the results for some well known interaction 
scenarios, namely $Q = 3H \xi \rho_x$, $Q = 3H \xi \rho_c$ and $Q = 3 H \xi 
(\rho_c+\rho_x)$.} }
\label{fig:coincidence-1}
\end{figure} 
\begin{figure*}
\includegraphics[width=0.4\textwidth]{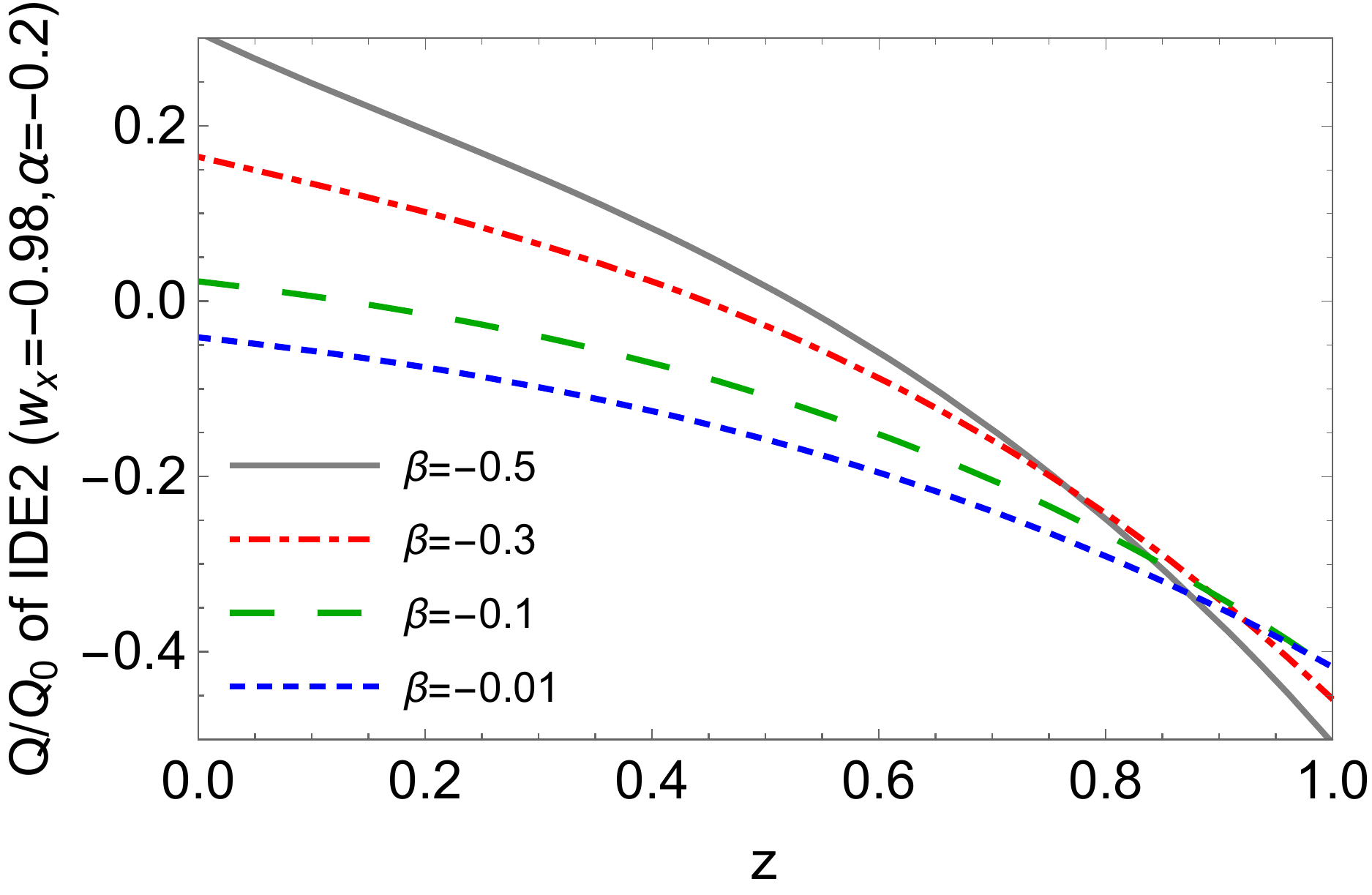}
\includegraphics[width=0.4\textwidth]{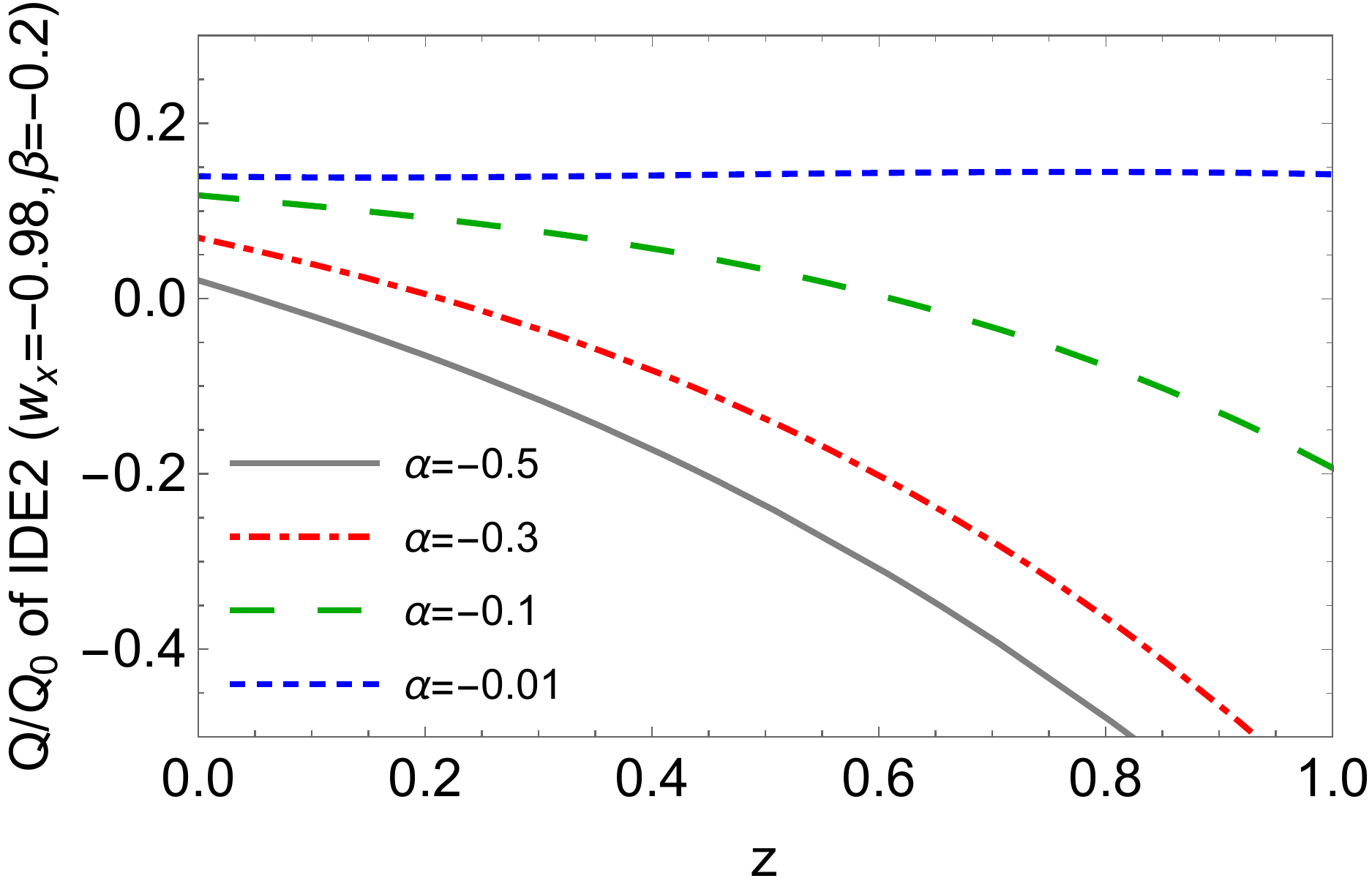}
\includegraphics[width=0.4\textwidth]{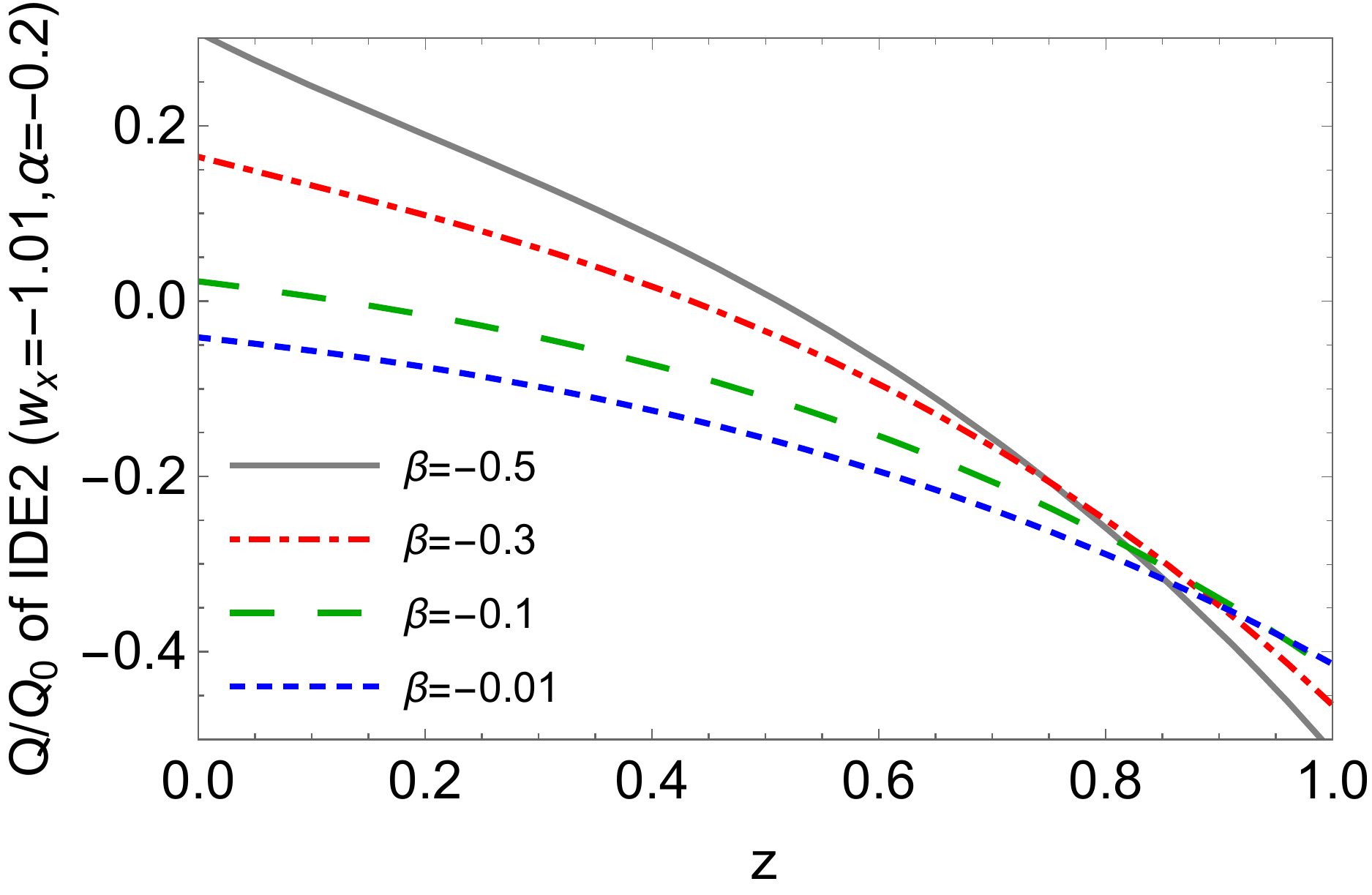}
\includegraphics[width=0.4\textwidth]{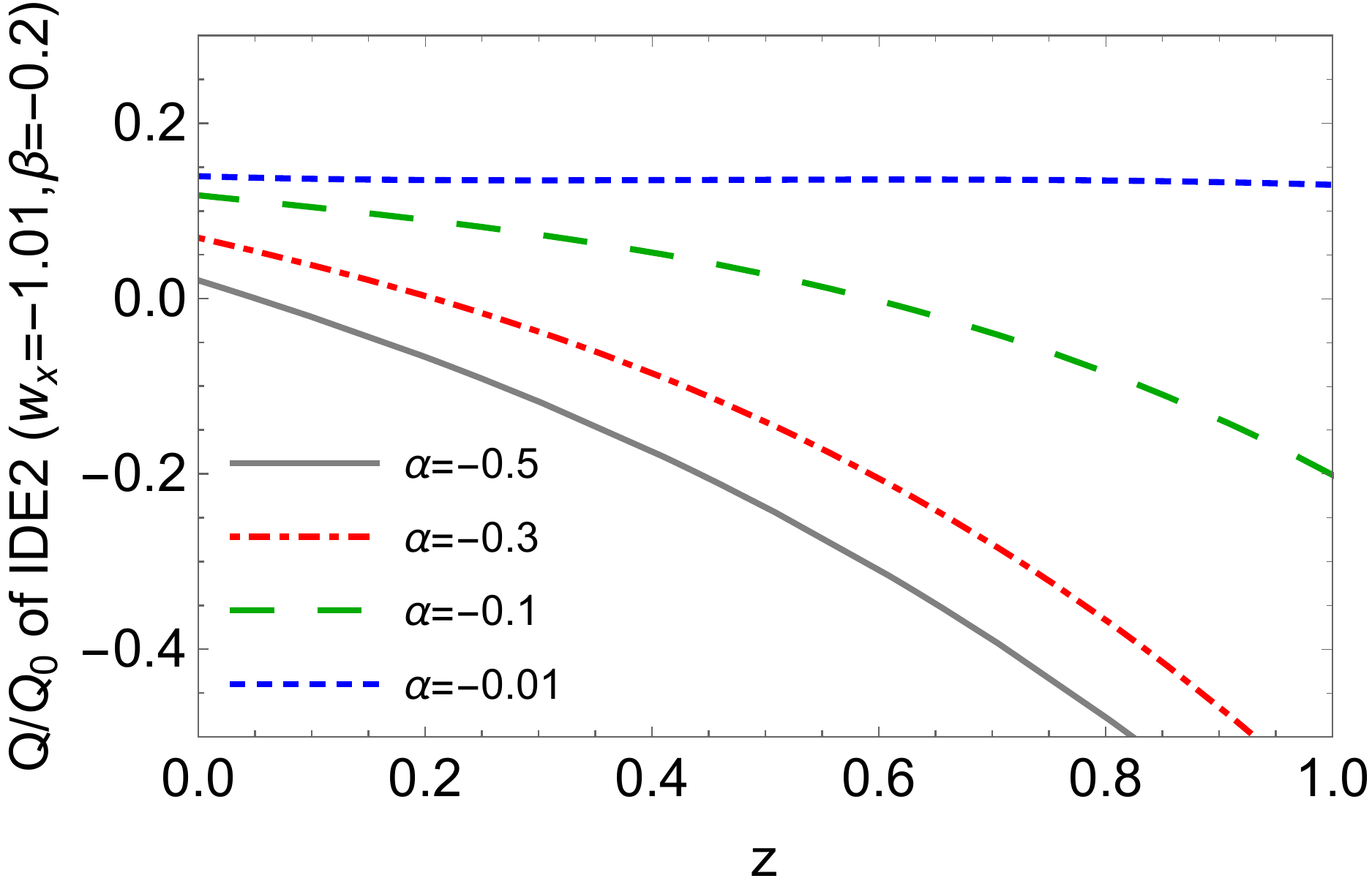}
\caption{{\it{
The evolution of the normalized interaction function $Q/Q_0$, where
$Q_0 = H_0 
\rho_{t0}$ (with $H_0$ and $\rho_{t0}$ the present values of the Hubble parameter and the 
total energy density 
$\rho_t$ respectively),
for the interaction model IDE2 of (\ref{IDE2}), for 
various values of the coupling parameters  $\alpha$ and $\beta$ in units where $8\pi 
G=1$, 
 for $w_x  
=-0.98$ (upper graphs) and for  $w_x = 
-1.01$ (lower graphs).}}}
\label{fig:Q-IDE2}
\end{figure*}

In  Fig. \ref{fig:q-IDE1} we depict the evolution of the deceleration
parameter for model IDE1, for $w_x > -1$ and $w_x < -1$, and choosing various values 
of the coupling parameter $\xi$. As we observe, the value of $\xi$ may make faster or 
delay the transition from deceleration to acceleration, which is a significant advantage 
in 
fitting observations precisely. Additionally, note that  the 
qualitative features  do not change for either $w_x> -1$ or 
$w_x < -1$. A similar graph can be obtained in the case of model IVS1, namely when $w_x= 
-1$.

We mention that in this scenario one can extract the solution for the ratio 
$\frac{\rho_x}{\rho_c}$   analytically. In particular, inserting (\ref{IDE1}) into  
(\ref{cons-cdm}) and (\ref{cons-de}) gives 
\begin{equation}
\frac{\rho_x}{\rho_c}=-\frac{w_x}{2\xi}-\sqrt{1+\frac{w_x^2}{4 
\xi^2}}\left(\frac{1-ba^{3\sqrt{4 \xi^2+  w_x^2}}}{1+ba^{3\sqrt{4 \xi^2+  w_x^2}}}\right),
\end{equation}
where 
\begin{equation}
b\equiv-\frac{2\xi\frac{\Omega_{x0}}{\Omega_{c0}}+w_x+\sqrt{4 \xi^2+  
w_x^2}}{2\xi\frac{\Omega_{x0}}{\Omega_{c0}}+w_x-\sqrt{4 \xi^2+  w_x^2}}.
\end{equation}

We proceed by investigating the coincidence parameter $r  = 
\rho_c/\rho_x$ for 
this interaction model. As it is known, alleviation of the coincidence problem requires a 
non-zero value for $r$ at late times. The evolution of the coincidence parameter for the model IDE1 is 
presented in Fig. \ref{fig:coincidence-1}, for different dark energy equation of 
states and setting  a typical 
value of $\xi = 0.01$ (for different values of $\xi$ the qualitative nature of 
the curves remains unaffected). Moreover, for comparison we also depict the 
results for some well known interaction scenarios of the literature, namely $Q = 3H \xi 
\rho_x$, $Q = 3H \xi \rho_c$ and $Q = 3 H \xi 
(\rho_c+\rho_x)$.
As one can see,   as $z \rightarrow 0$ all curves 
tend to  a non-zero value, which implies that the coincidence problem is alleviated. 
Additionally, although at low redshifts all models present the same behavior,   
at high redshifts the sign-changeable interaction models considered in this work are 
distinguishable from the known interacting models of the literature.

 Finally, concerning the perturbations in this scenario, they are determined by  
(\ref{perteq1})-(\ref{perteq4}), with 
\begin{eqnarray}  
 \frac{\delta 
Q}{Q}=\frac{\delta_c\rho_c-\delta_x\rho_x}{\rho_c-\rho_x}+\frac{2\theta+h'}{6 
\mathcal{H}},
\end{eqnarray}%
 with $\theta\equiv\theta_\mu^\mu$ the volume expansion of the total fluid.
 
  \begin{figure*}
\includegraphics[width=0.4\textwidth]{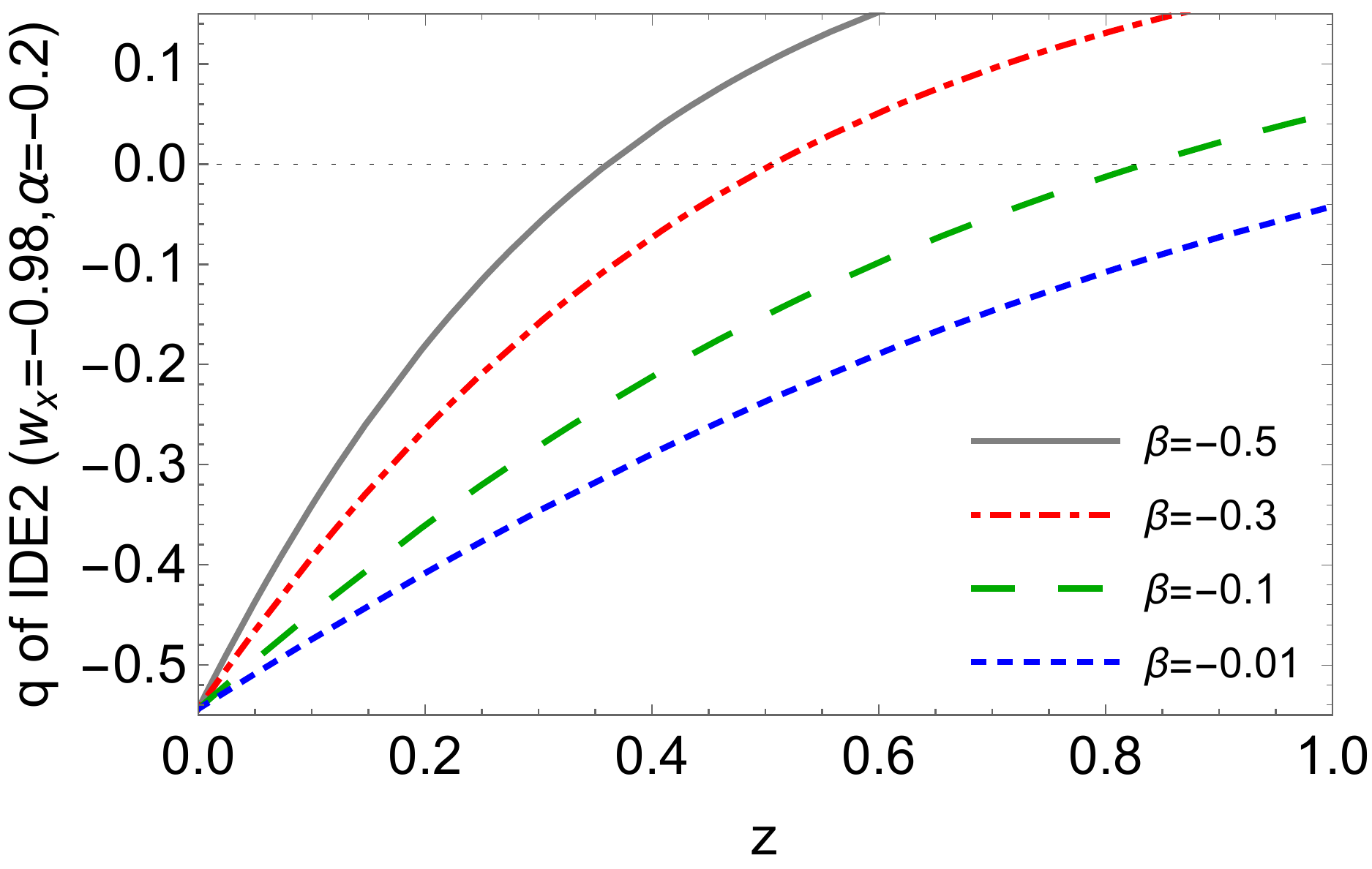}
\includegraphics[width=0.4\textwidth]{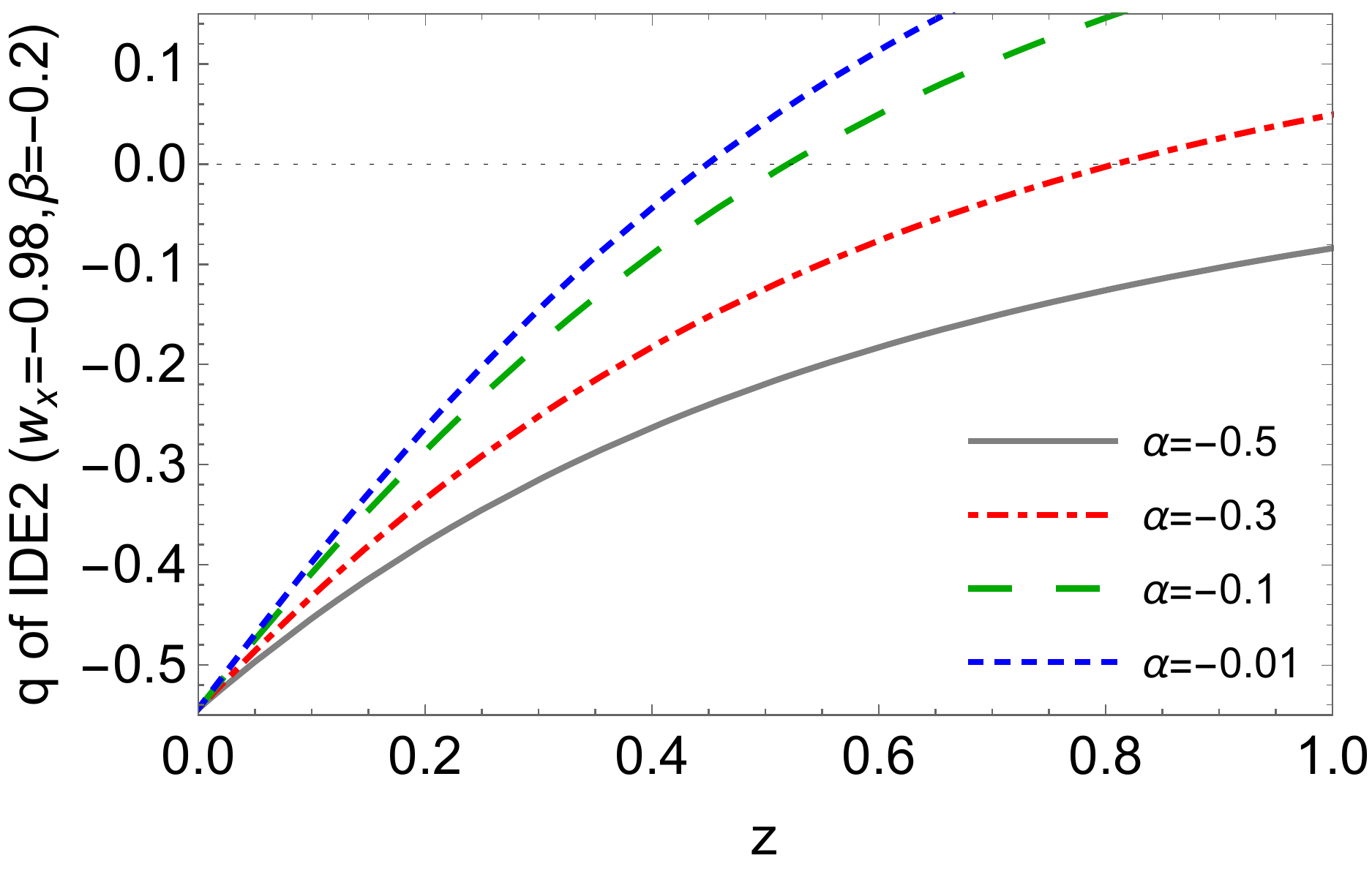}
\includegraphics[width=0.4\textwidth]{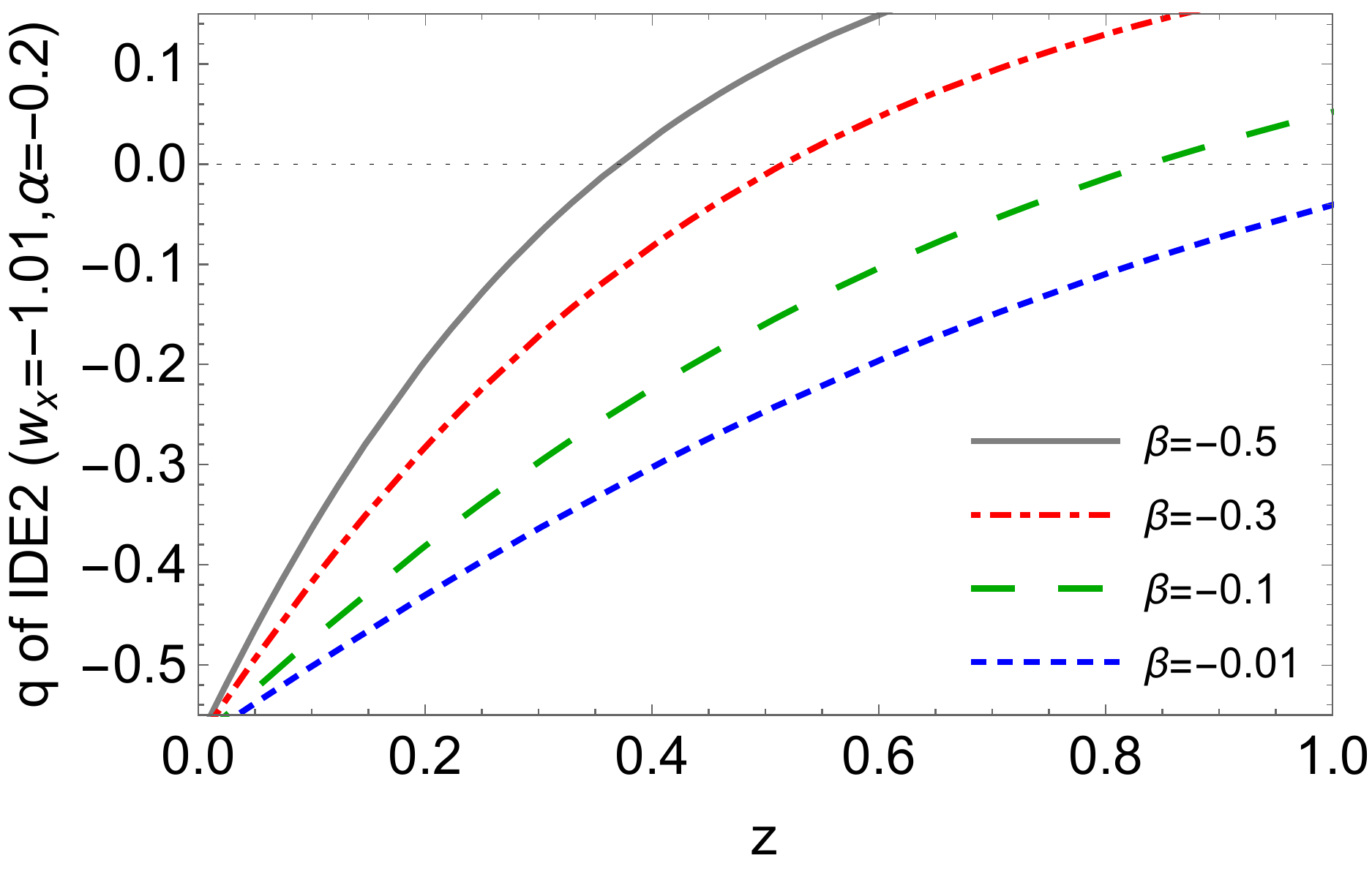}
\includegraphics[width=0.4\textwidth]{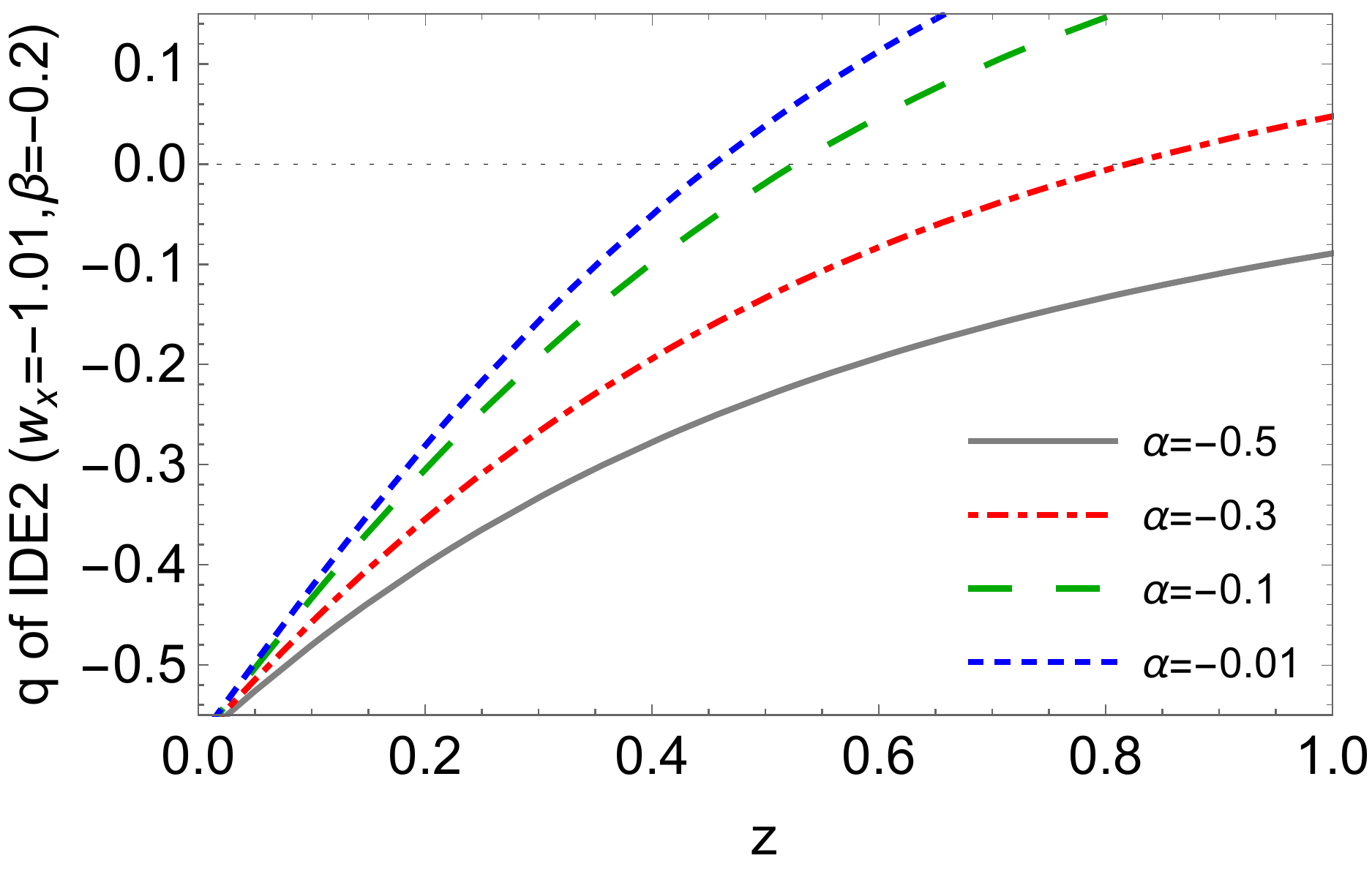}
\caption{{\it{The evolution of the deceleration parameter as a function of the redshift, 
for the interaction model IDE2 of (\ref{IDE2}), for $w_x  =-0.98$ (upper graphs) and for  
$w_x = 
-1.01$ (lower graphs), for 
various values of the coupling parameters $\alpha$ and $\beta$  in units where $8\pi 
G=1$. We have set
 $\Omega_{x0}\approx0.69$, $\Omega_{c0}\approx0.25$, 
$\Omega_{b0}\approx0.05$ and $\Omega_{r0}\approx10^{-4}$ in agreement with 
observations.}}}
\label{fig:q-IDE2}
\end{figure*}
 
\begin{figure*}
\includegraphics[width=0.39\textwidth]{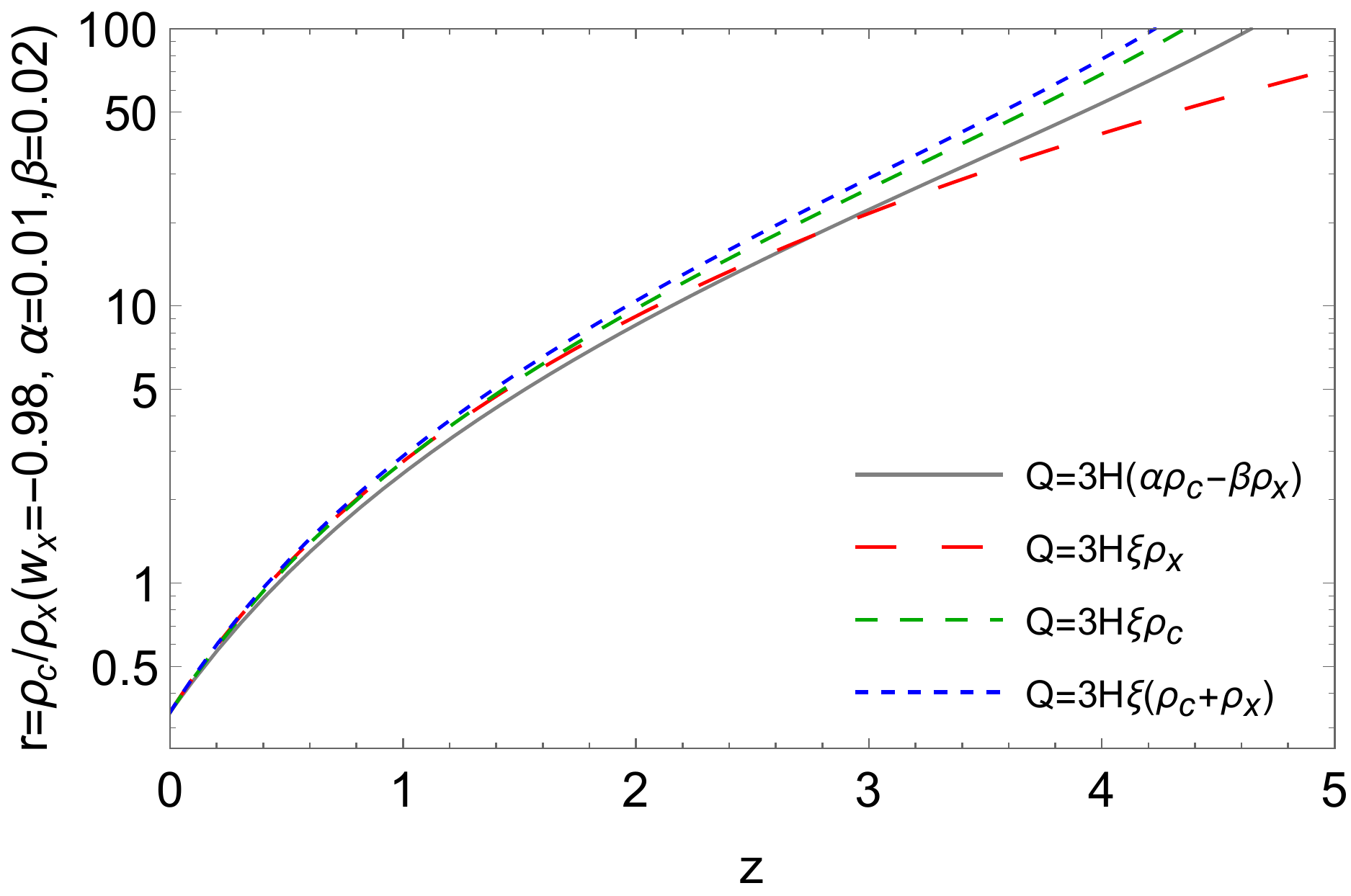}
\includegraphics[width=0.39\textwidth]{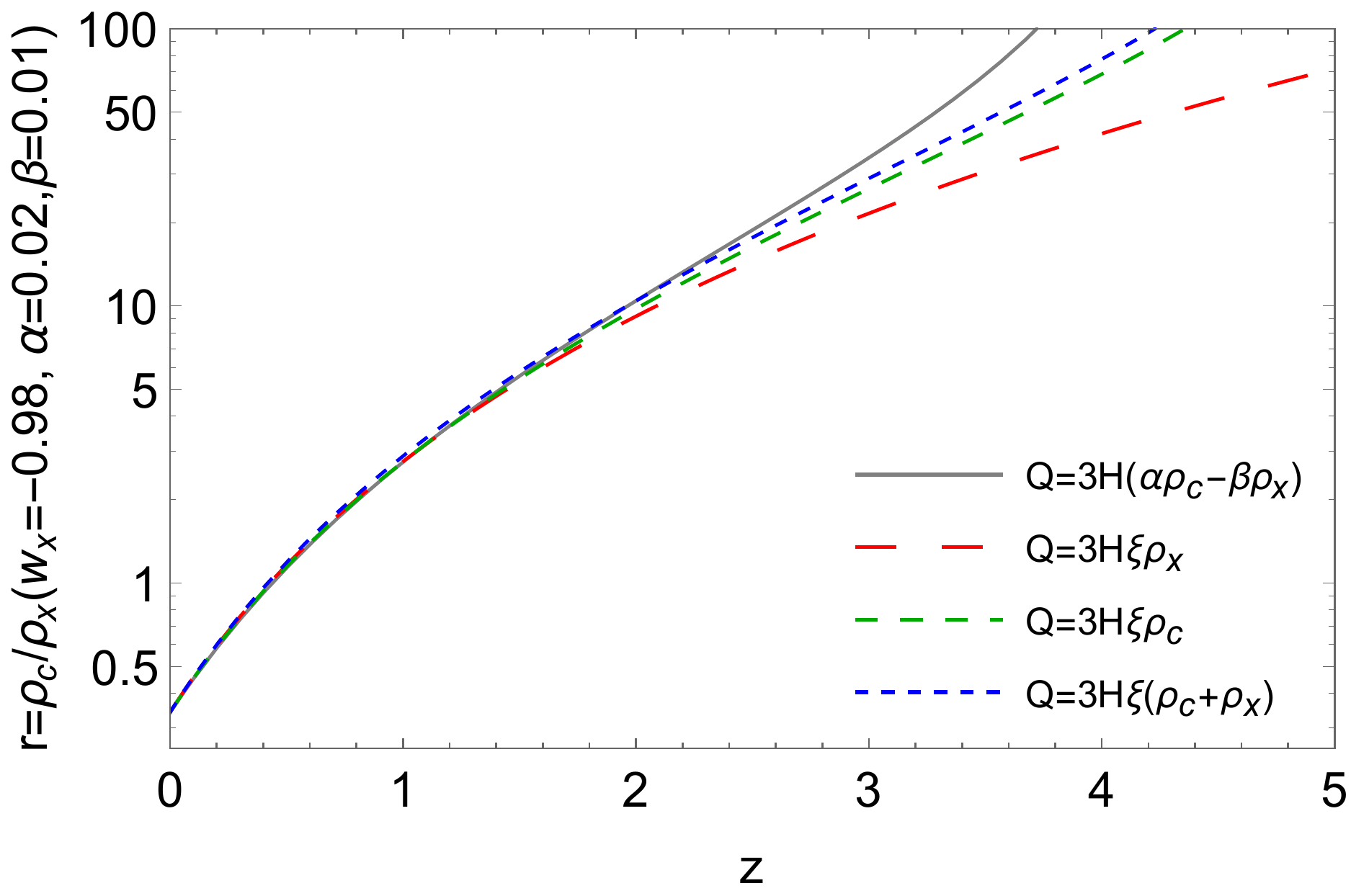}\\
\includegraphics[width=0.39\textwidth]{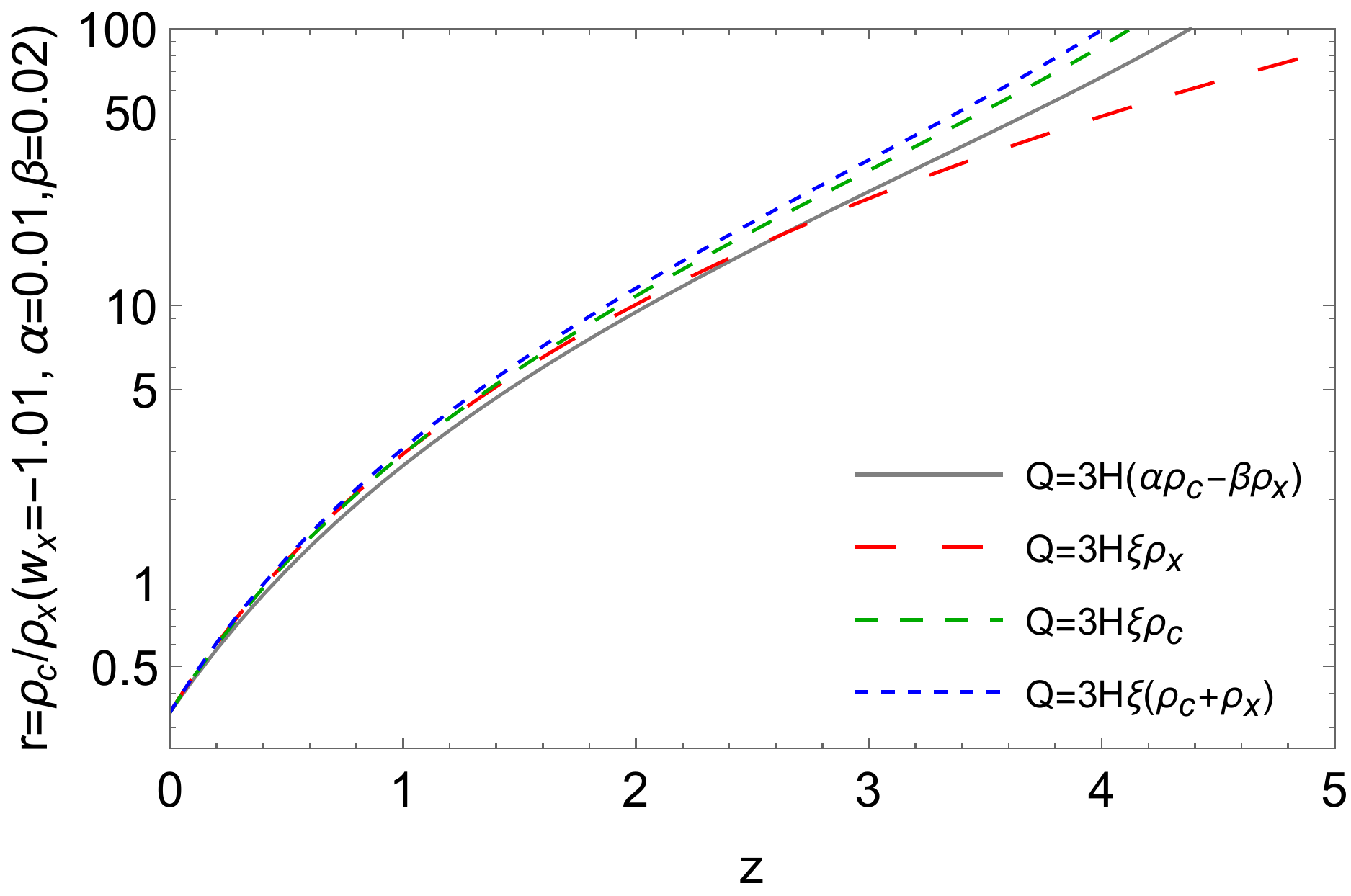}
\includegraphics[width=0.39\textwidth]{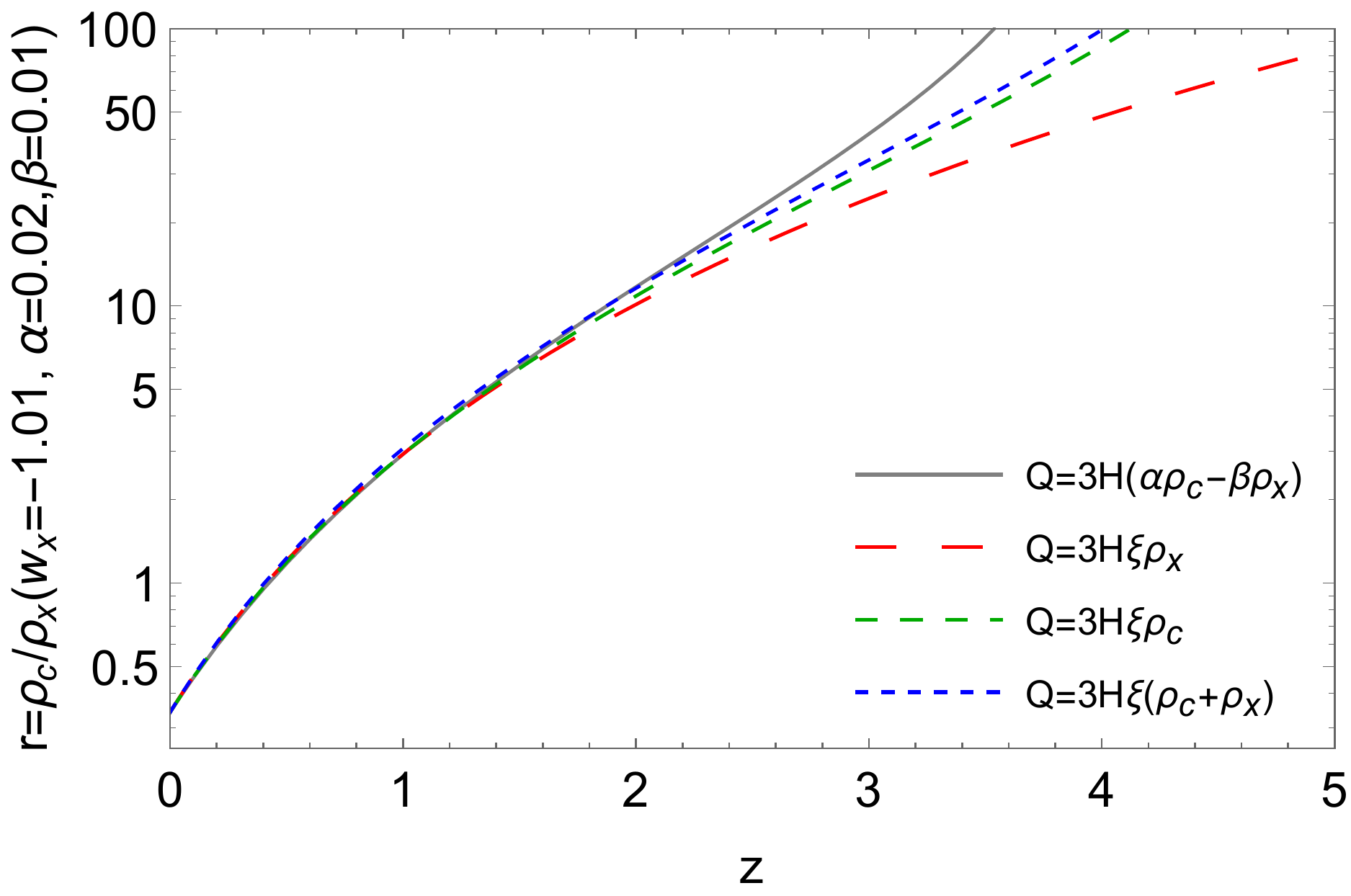}
\caption{\it{
The evolution of the coincidence parameter $r = \rho_c/\rho_x$,
for the interaction model IDE2 of (\ref{IDE2}),  for 
various values of the coupling parameters  $\alpha$ and $\beta$ in units where $8\pi 
G=1$,  for $w_x  
=-0.98$ (upper graphs) and for  $w_x = 
-1.01$ (lower graphs).
 For comparison we additionally depict the results for some well known interaction 
scenarios, namely $Q = 3H \xi \rho_x$, $Q = 3H \xi \rho_c$ and $Q = 3 H \xi 
(\rho_c+\rho_x)$, taking  a 
typical value $\xi = 0.01$ in units where $8\pi 
G=1$.} } 
\label{fig:coincidence-2}
\end{figure*}

 \subsection{Interaction function  $Q = 3 H (\alpha \rho_c - \beta \rho_x)$}

 We proceed with the investigation of the interacting model
\begin{eqnarray}\label{IDE2}
Q  = 3 H (\alpha \rho_c - \beta \rho_x), 
\end{eqnarray}%
where $\alpha $, $\beta $ are the coupling parameters considered to have the same sign. 
In the case where $\alpha = \beta = \xi$ the above model coincides with the one of the 
previous subsection, namely (\ref{IDE1}). We consider two choices for 
 $w_x$, namely  $w_x= -1$ (from now on model IVS2), and the case $w_
x \neq -1$ (from now on model IDE2), which will be further  divided 
into $w_x > -1$ and $w_x < -1$.  
 
In Fig. \ref{fig:Q-IDE2} we   depict the evolution of $Q$ normalized by $Q_0 = H_0 
\rho_{t0}$, for various values of   the coupling 
parameters $\alpha$ and $\beta$.  Additionally, in Fig. \ref{fig:q-IDE2} we present the 
evolution of the deceleration
parameter for   model IDE2, for $w_x > -1$ and $w_x < -1$ and choosing various values 
of   $\alpha$ and $\beta$. As we observe, the values of the 
coupling parameters significantly determine the exact transition redshift from 
deceleration to 
acceleration, while the exact value of  $w_x$ does not have a significant effect.
 Similar graphs can be obtained in the case of  IVS2 model, namely when $w_x= 
-1$.

Inserting (\ref{IDE2}) into  
(\ref{cons-cdm}) and (\ref{cons-de}) provides as the analytical solution for the ratio 
$\frac{\rho_x}{\rho_c}$, namely
\begin{eqnarray}
&&
\!\!\!\!\!\!\!\!\!\!\!\!\!\!\!\!\!\!\!\!\!\! 
\frac{\rho_x}{\rho_c}=\frac{\alpha-\beta-w_x}{2 \beta}-\frac{\sqrt{
(\alpha-\beta-w_x)^2+4 
\alpha \beta}}{2 \beta}\nonumber\\
&& \ \ \ \ \ \ \   \ \ \   \ \ \   \ \ \  
\cdot
\left(\frac{1-ba^{3\sqrt{(\alpha-\beta-w_x)^2+4 \alpha 
\beta}}}{1+ba^{3\sqrt{(\alpha-
\beta-w_x)^2+4 \alpha \beta}}}\right),
\end{eqnarray}
with
\begin{equation}
b\equiv - \frac{2\beta\frac{\Omega_{x0}}{\Omega_{c0}}-(\alpha-\beta-w_x)+\sqrt{
(\alpha-\beta-w_x)^2+4 \alpha 
\beta}}{2\beta\frac{\Omega_{x0}}{\Omega_{c0}}-(\alpha-\beta-w_x)-\sqrt{
(\alpha-\beta-w_x)^2+4 
\alpha \beta}}.
\end{equation}
 
We now investigate the evolution of the coincidence parameter $r  = \rho_c/\rho_x$ 
for this 
interaction model. In Fig. \ref{fig:coincidence-2}, we depict the evolution of $r$
for 
various values of the coupling parameters  $\alpha$ and $\beta$ and  
for different dark energy equation of 
states. Furthermore, for comparison
we also depict the 
results for some well known interaction scenarios, namely $Q = 3H \xi \rho_x$, $Q = 3H 
\xi 
\rho_c$ and $Q = 3 H \xi 
(\rho_c+\rho_x)$.
Similarly to model IDE1 above, for the present model IDE2 we can see that for $z 
\rightarrow 0$ all curves 
tend to  a non-zero value, which implies that the coincidence problem is alleviated. 
Additionally, although at low redshifts all models present similar behavior,   
at high redshifts the sign-changeable interaction models considered here are 
distinguishable from the known interacting models of the literature.

 Lastly, concerning the perturbations, in this scenario they are determined by  
(\ref{perteq1})-(\ref{perteq4}), with 
\begin{eqnarray}
\frac{\delta Q}{Q}= 
\frac{\alpha\delta_c\rho_c-\beta\delta_x\rho_x}{\alpha\rho_c-\beta\rho_x}+\frac{
2\theta +h^{\prime }}{6\mathcal{H}}.
\end{eqnarray}%

\section{The data and methodology}
\label{sec-data}

In this section we briefly  describe the observational datasets and the methodology we 
follow in order to  constrain the aforementioned interaction models.

\begin{enumerate}

\item Cosmic microwave background (CMB) radiation: We consider the CMB  data from Planck 
2015 measurements \cite{Adam:2015rua,Aghanim:2015xee}, and in particular  we use 
the high-$\ell$  and   low-$\ell$  temperature and polarization 
   data from  \cite{Adam:2015rua,Aghanim:2015xee}.

\item  Baryon acoustic oscillation (BAO): We use data from  BAO distance  
measurements from the following sources. Data from 6dF
Galaxy Survey (6dFGS) (redshift measurement at $z_{\emph{\emph{eff}}}=0.106$%
) \cite{Beutler:2011hx}, data from Main Galaxy Sample of Data Release 7 of Sloan
Digital Sky Survey (SDSS-MGS) ($z_{\emph{\emph{eff}}}=0.15$) \cite%
{Ross:2014qpa}, and   data from CMASS and LOWZ samples of the latest Data Release 12
(DR12) of the Baryon Oscillation Spectroscopic Survey (BOSS) ($z_{\mathrm{eff%
}}=0.57$) \cite{Gil-Marin:2015nqa} and ($z_{\mathrm{eff}}=0.32$) \cite%
{Gil-Marin:2015nqa}.

\item Supernovae Type Ia (SNIa): We use the most latest compilation of SNIa,
consisting of $1048$ data points spanned over the redshift interval $z \in [0.01, 2.3]$ 
known as the Pantheon sample \cite{Scolnic:2017caz}.

\item Cosmic Chronometers (CC): We use the Hubble parameter measurements from the  cosmic 
chronometers. The total number of data is 30 and the measurements are spanned over the 
redshift 
interval $0< 
z < 2$ \cite{Moresco:2016mzx}. For technical details we further refer to 
\cite{Moresco:2016mzx}. 
 
\end{enumerate}

In order to perform the analysis and extract the observational constraints, we   
use the Markov chain Monte Carlo package \texttt{\small COSMOMC} \cite{Lewis:2002ah,
Lewis:1999bs} where a convergence diagnostic by Gelman-Rubin is included 
\cite{Gelman-Rubin}, which 
in addition supports the
Planck 2015 likelihood code \cite{Aghanim:2015xee}\footnote{See the publicly available 
code  at \url{http://cosmologist.info/cosmomc/}. Note that although 
Planck 
2018 cosmological parameters \cite{Aghanim:2018eyx} are now available, the Planck 2018 
likelihoods  are still not available publicly, and thus we use the   Planck 2015 
ones. }.
In the case of IDE1 model  we have the eight-dimensional parameter space: 
$\mathcal{P}_1 \equiv\Bigl\{\Omega_bh^2, \Omega_{c}h^2, 100 
\theta_{MC}, \tau, w_x, n_s, log[10^{10}A_S], \xi\Bigr\}$,   while  
for IVS1 we have one parameter less since $w_x  =-1$.  
For IDE2  we have the 
nine-dimensional parameter space $\mathcal{P}_2 \equiv \Bigl\{\Omega_bh^2, 
\Omega_{c}h^2, 100 \theta_{MC},\tau, w_x, n_s, log[10^{10}A_S],\alpha, \beta\Bigr\}$, 
and 
similarly for IVS2 we have one parameter less. In the above expressions  
 $\Omega_bh^2$ is the physical baryons density, $\Omega_{c}
h^2$ is the cold dark matter density, $100 \theta_{MC}$ is the ratio of sound horizon to 
the angular diameter distance, $\tau$ is the
optical depth, $w_x$ is the equation-of-state parameter for dark energy, $n_s$ is the 
scalar spectral index, and $A_S$ is the amplitude of the initial power spectrum.   
The 
remaining parameters in $\mathcal{P}_1$ and $\mathcal{P}_2$, namely $\xi$, $\alpha$ and
$\beta$ are 
the coupling parameters for the interaction models. 
Finally, in Table \ref{tab:priors-ide12} we present the flat priors imposed on the 
free parameters of the prescribed interacting scenarios.

\begin{table}[ht]
\begin{center}
\begin{tabular}{|c|c|c|}
\hline
Parameter                    & Prior \\
\hline 
$\Omega_{b} h^2$             & $[0.005,0.1]$\\
$\Omega_{c} h^2$             & $[0.01,0.99]$\\
$\tau$                       & $[0.01,0.8]$\\
$n_s$                        & $[0.5, 1.5]$\\
$\log[10^{10}A_{s}]$         & $[2.4,4]$\\
$100\theta_{MC}$             & $[0.5,10]$\\ 
$w_x$                        & $[-2, 0]$\\
$\xi$                        & $[-1, 0]$\\ 
$\alpha $                    & $[-1, 0]$ \\
$\beta $                     & $[-1, 0]$ \\
\hline 
\end{tabular}
\end{center}
\caption{The flat priors on the cosmological parameters used in the present analyses. }
\label{tab:priors-ide12}
\end{table}

\begin{center}                       
\begin{table*}             
\begin{tabular}{ccccccccccccccc}         
\hline\hline                
Parameters & CMB+BAO &  CMB+BAO+Pantheon & CMB+BAO+Pantheon+CC \\ \hline
$\Omega_c h^2$ & $    0.1198_{-    0.0013-    0.0026}^{+    0.0013+    0.0024}$ & $ 
0.1196_{-    0.
0012-    0.0023}^{+    0.0012+    0.0024}$ &  $    0.1196_{-    0.0013-    0.0023}^{+    
0.0011+    
0.0024}$ & \\
$\Omega_b h^2$ & $    0.02231_{-    0.00016-    0.00029}^{+    0.00016+    0.00032}$ & $  
  0.02231_
{-    0.00015-    0.00031}^{+    0.00016+    0.00029}$ & $    0.02230_{-    0.00016-    
0.00029}^{+ 
   0.00015+    0.00030}$ &\\
$100\theta_{MC}$ & $    1.04051_{-    0.00031-    0.00066}^{+    0.00032+    0.00070}$ & 
$    1.
04054_{-    0.00032-    0.00064}^{+    0.00033+    0.00064}$ & $    1.04053_{-    
0.00030- 
   0.
00058}^{+    0.00031+    0.00057}$ &\\
$\tau$ & $    0.082_{-    0.017-    0.032}^{+    0.017+    0.032}$ & $    0.084_{-    
0.016-    0.
031}^{+    0.017+    0.030}$ & $    0.082_{-    0.017-    0.034}^{+    0.016+    0.032}$ 
&\\
$n_s$ & $    0.9741_{-    0.0041-    0.0080}^{+    0.0041+    0.0078}$ & $    0.9741_{-   
 0.0046-  
  0.0080}^{+    0.0040+    0.0080}$ & $    0.9742_{-    0.0038-    0.0074}^{+    0.0037+  
 
 0.0073}
$ & \\
${\rm{ln}}(10^{10} A_s)$ & $    3.104_{-    0.034-    0.063}^{+    0.034+    0.066}$ &  $ 
  3.108_{
-    0.031-    0.061}^{+    0.033+    0.065}$ & $    3.105_{-    0.033-    0.066}^{+    
0.032+    0.
063}$ &\\
$w_x$ & $   -1.0818_{-    0.0328-    0.0916}^{+    0.0634+    0.0807}$ & $   -1.0529_{-   
0.0206-  
  0.0555}^{+    0.0352+    0.0483}$ & $   -1.0545_{-    0.0204-    0.0545}^{+    0.0342+  
 
 0.0482}
$ &\\
$\xi$ & $   -0.00014_{-    0.00004-    0.00019}^{+    0.00014+    0.00014}$ & $   
-0.00013_{-    0.
00003-    0.00020}^{+  0.00013+  0.00013}$ & $   -0.00013_{-    0.00003-    0.00020}^{+   
 
0.00013+ 
   0.00013}$ &\\
$\Omega_{m0}$ & $    0.299_{-    0.010-    0.020}^{+    0.011+    0.018}$ & $    0.305_{- 
   0.008- 
   0.013}^{+    0.007+    0.014}$ & $    0.305_{-    0.007-    0.014}^{+    0.007+    
0.014}$ & \\
$\sigma_8$ & $    0.845_{-    0.017-    0.034}^{+    0.017+    0.036}$ & $    0.840_{-    
0.014-    
0.027}^{+    0.015+    0.027}$ & $    0.839_{-    0.014-    0.028}^{+    0.014+    
0.030}$ 
&\\
$H_0$ & $   69.12_{-    1.39-    2.11}^{+    0.93+    2.40}$ & $   68.39_{-    0.69-    
1.37}^{+    
0.68+    1.39}$ & $   68.43_{-    0.76-    1.39}^{+    0.67+    1.44}$ &\\
\hline\hline                                                    
\end{tabular}                                
\caption{Summary of the 68\% ($1 \sigma$) and 95\% ($ 2 \sigma$) confidence-level (CL) 
constraints on  the interaction model IDE1 of (\ref{IDE1}),
using various combinations of the observational 
data sets. Here, $\Omega_{m0}$ denotes the present value of $\Omega_m = 
\Omega_b+\Omega_c$ and $H_0$ is in the units of km/s/Mpc.  }
\label{tab:constraints-Model1}             
\end{table*}                
\end{center}

\begin{figure*}
\includegraphics[width=0.6\textwidth]{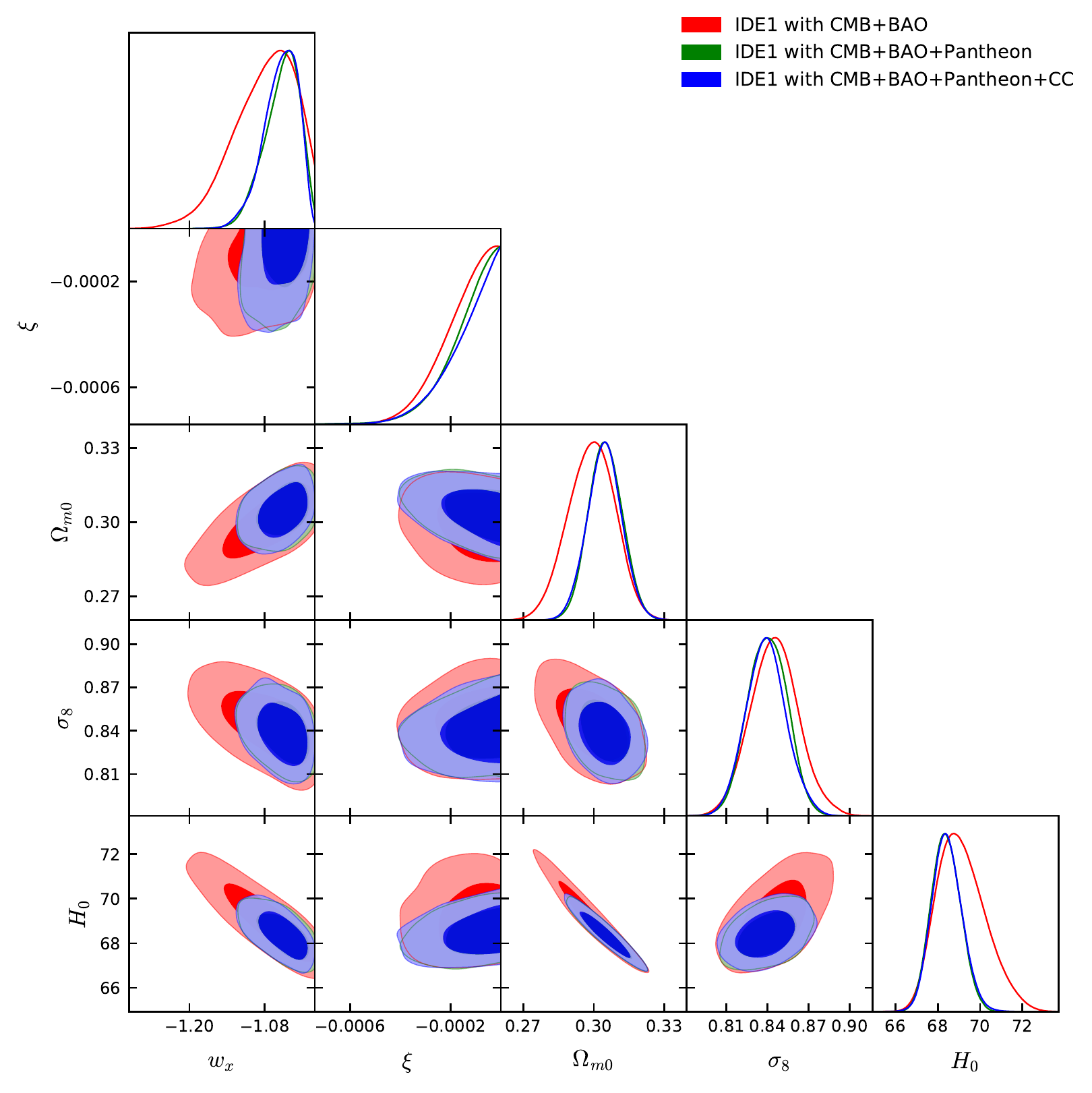}
\caption{{\it{The  $1 \sigma$ and $ 2 \sigma$ CL   contour 
plots for several combinations of various quantities and using
various combinations of the observational data sets, 
for the interaction model IDE1 of (\ref{IDE1}), and the corresponding 1-dimensional (1D) 
marginalized posterior distributions.
 Here, $\Omega_{m0}$ denotes the present value of $\Omega_m = 
\Omega_b+\Omega_c$ and $H_0$ is in the units of km/s/Mpc.  }}}
\label{contour:ide1}
\end{figure*}

\begin{center}                 
\begin{table*}                
\begin{tabular}{cccccccccccccc}           
\hline\hline                
Parameters & CMB+BAO & CMB+BAO+Pantheon & CMB+BAO+Pantheon+CC  \\ \hline
$\Omega_c h^2$ & $    0.1187_{-    0.0010-    0.0020}^{+    0.0011+    0.0020}$ & $    
0.1184_{-    
0.0010-    0.0019}^{+    0.0010+    0.0019}$ & $    0.1184_{-    0.0009-    0.0020}^{+    
0.0009+   
 0.0019}$\\
$\Omega_b h^2$ & $    0.02234_{-    0.00015-    0.00028}^{+    0.00014+    0.00029}$  & $ 
   0.
02236_{-    0.00015-    0.00028}^{+    0.00015+    0.00028}$ & $    0.02235_{-    
0.00015- 
   0.
00030}^{+    0.00014+    0.00030}$\\
$100\theta_{MC}$ & $    1.04064_{-    0.00030-    0.00059}^{+    0.00030+    0.00060}$  & 
$    1.
04066_{-    0.00035-    0.00059}^{+    0.00030+    0.00067}$ & $    1.04066_{-    
0.00032- 
   0.
00060}^{+    0.00032+    0.00061}$ \\

$\tau$ & $    0.088_{-    0.016-    0.033}^{+    0.016+    0.031}$ & $    0.089_{-    
0.017-    0.
032}^{+    0.017+    0.032}$ & $    0.087_{-    0.017-    0.031}^{+    0.016+    0.032}$\\

$n_s$ & $    0.9767_{-    0.0037-    0.0077}^{+    0.0038+    0.0077}$ & $    0.9773_{-   
 
0.0038-  
  0.0070}^{+    0.0035+    0.0070}$ & $    0.9773_{-    0.0038-    0.0072}^{+    0.0035+  
 
 0.0076}
$\\
${\rm{ln}}(10^{10} A_s)$ & $    3.115_{-    0.031-    0.064}^{+    0.032+    0.062}$  & $ 
 
  3.117_{
-    0.034-    0.065}^{+    0.033+    0.063}$ & $    3.112_{-    0.034-    0.061}^{+    
0.033+    0.
064}$ \\
$\xi$ & $   -0.000089_{-    0.000018-    0.000151}^{+    0.000089+    0.000089}$& $   
-0.000080_{-  
  0.000016-    0.000139}^{+    0.000080+    0.000080}$ & $   -0.000078_{-    0.000018-    
0.000129}
^{+    0.000078+    0.000078}$  \\
$\Omega_{m0}$ & $    0.312_{-    0.007-    0.013}^{+    0.007+    0.014}$  & $    
0.310_{- 
   0.006-
    0.013}^{+    0.006+    0.013}$ & $    0.311_{-    0.006-    0.012}^{+    0.006+    
0.012}$\\
$\sigma_8$ & $    0.825_{-    0.014-    0.027}^{+    0.014+    0.026}$ & $    0.825_{-    
0.014-    
0.028}^{+    0.014+    0.027}$ & $    0.824_{-    0.014-    0.026}^{+    0.014+    
0.026}$\\
$H_0$ & $   67.36_{-    0.52-    1.03}^{+    0.52+    1.02}$ & $   67.50_{-    0.48-    
0.98}^{+    
0.48+    0.97}$ & $   67.49_{-    0.46-    0.91}^{+    0.46+    0.95}$\\
\hline\hline                 
\end{tabular}                    
\caption{Summary of the  $1 \sigma$ and  $ 2 \sigma$  CL 
constraints on  the interaction model IVS1,
using various combinations of the observational 
data sets. Here, $\Omega_{m0}$ denotes the present value of $\Omega_m = 
\Omega_b+\Omega_c$ and $H_0$ is in the units of km/s/Mpc.
 }\label{tab:cons-ivs1}                                       
\end{table*}                     
\end{center}

\begin{figure}
\includegraphics[width=0.4\textwidth]{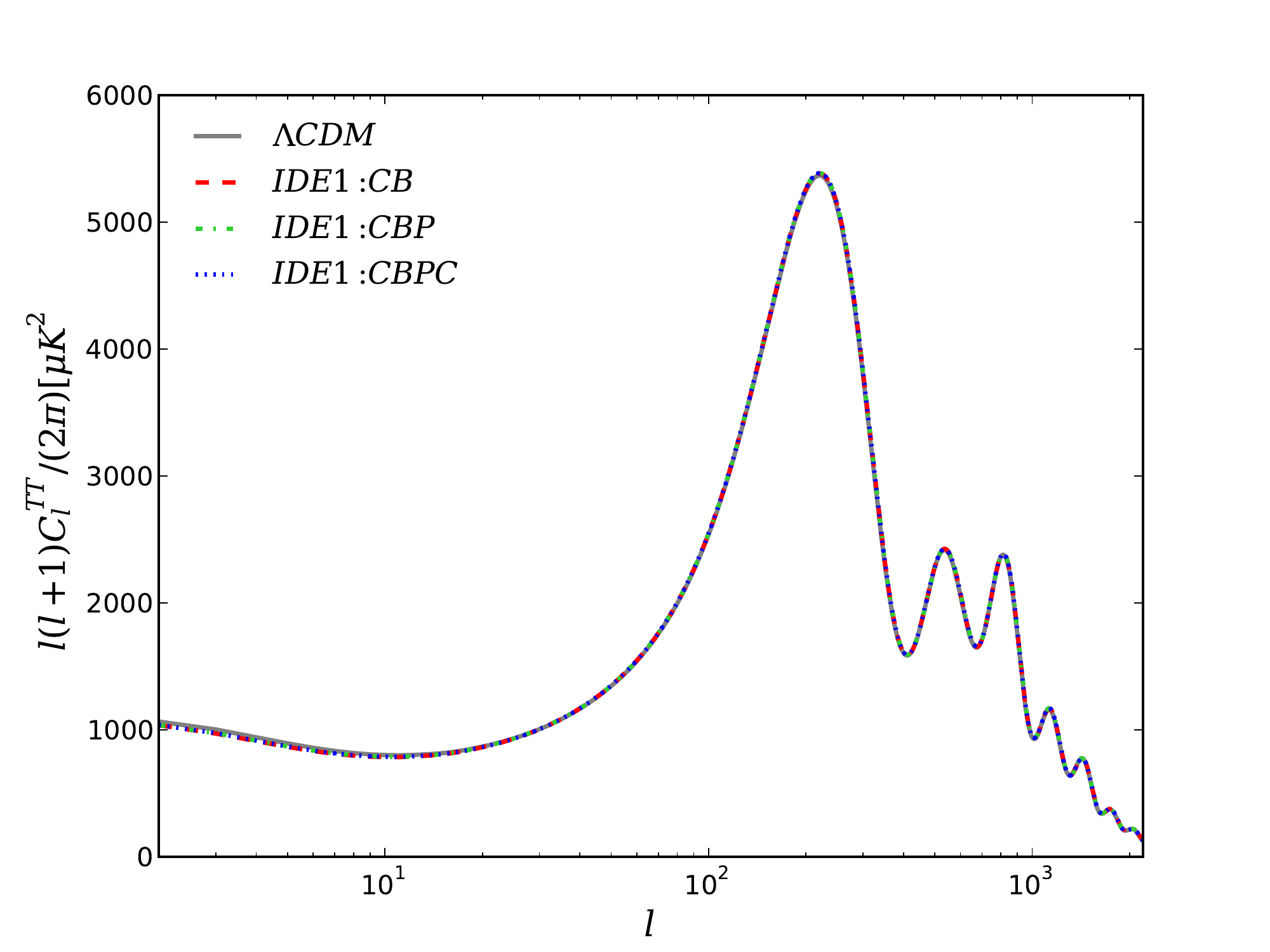}
\includegraphics[width=0.4\textwidth]{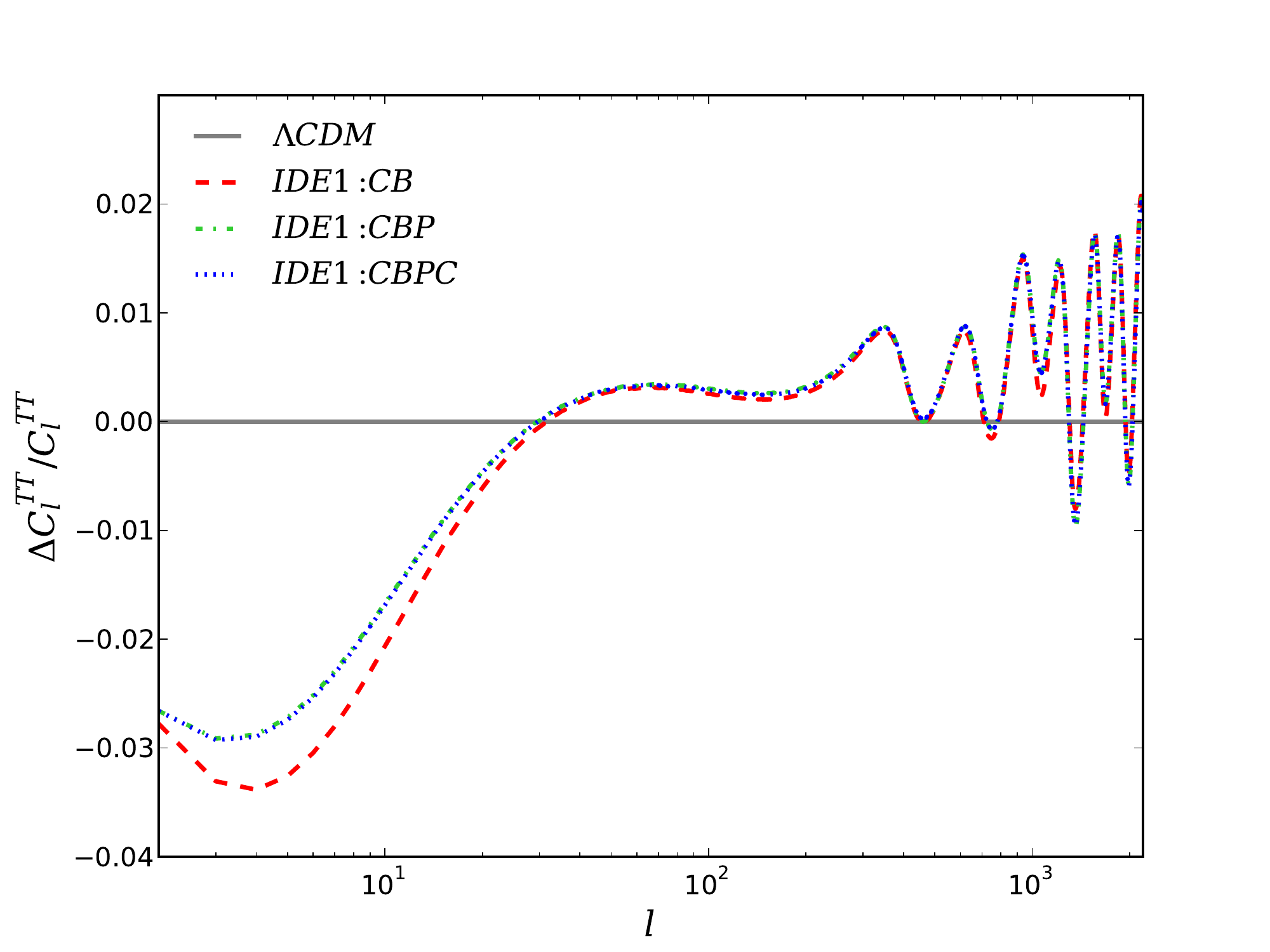}
\caption{{\it{The temperature anisotropy in the CMB TT spectra for the interaction 
model IDE1 of (\ref{IDE1}), considering three different combinations of the observational 
datasets, namely CB ($=$ CMB+BAO), CBP ($=$CMB+BAO+Pantheon)  and CBPC ($=$ 
CMB+BAO+Pantheon+CC), as well as the   curve for $\Lambda$CDM paradigm (upper 
graph), and the corresponding residual plot with reference to 
$\Lambda$CDM scenario (lower graph).}}}
\label{fig-cmb-ide1}
\end{figure}

\begin{figure}
\includegraphics[width=0.4\textwidth]{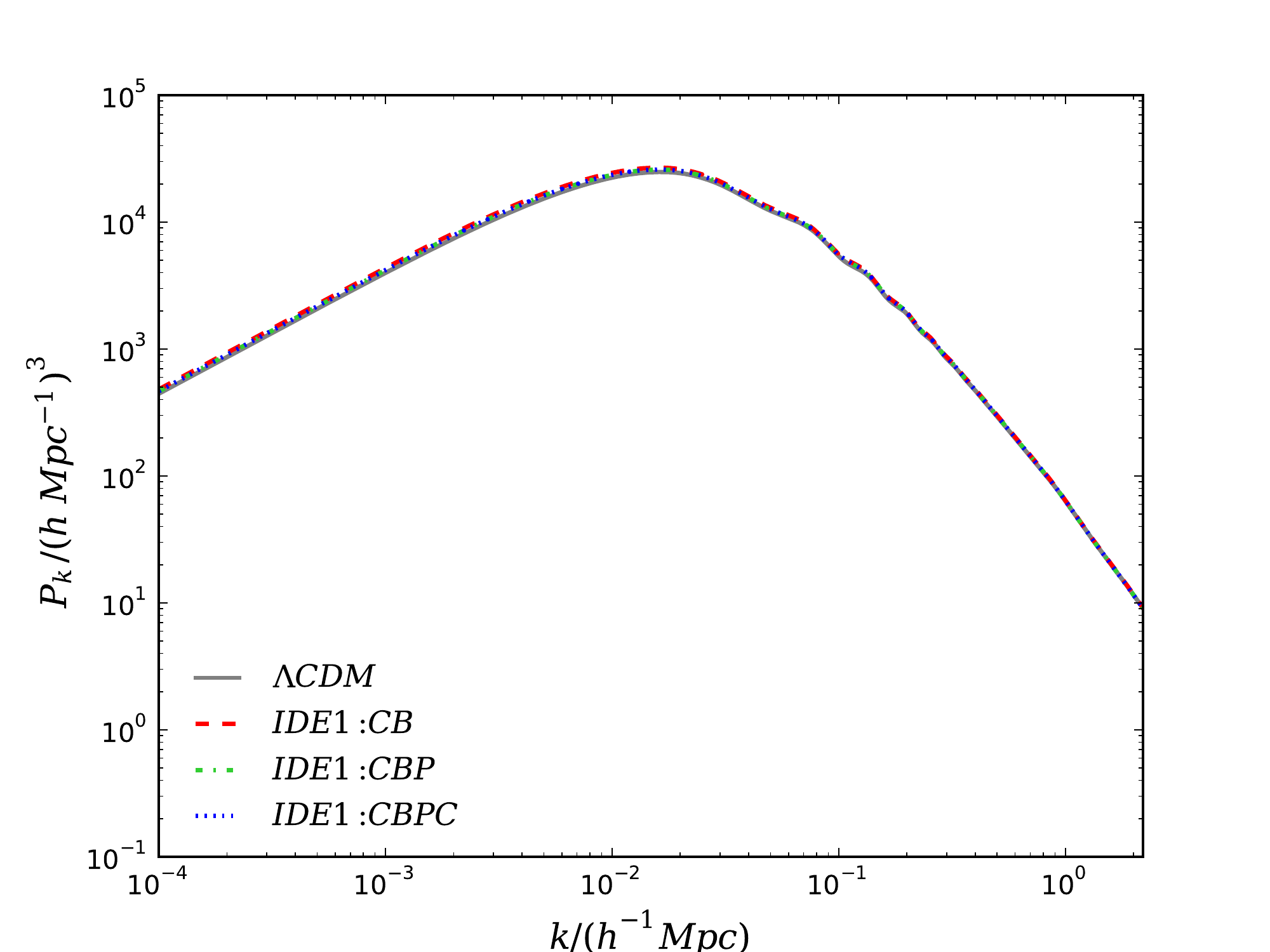}
\includegraphics[width=0.4\textwidth]{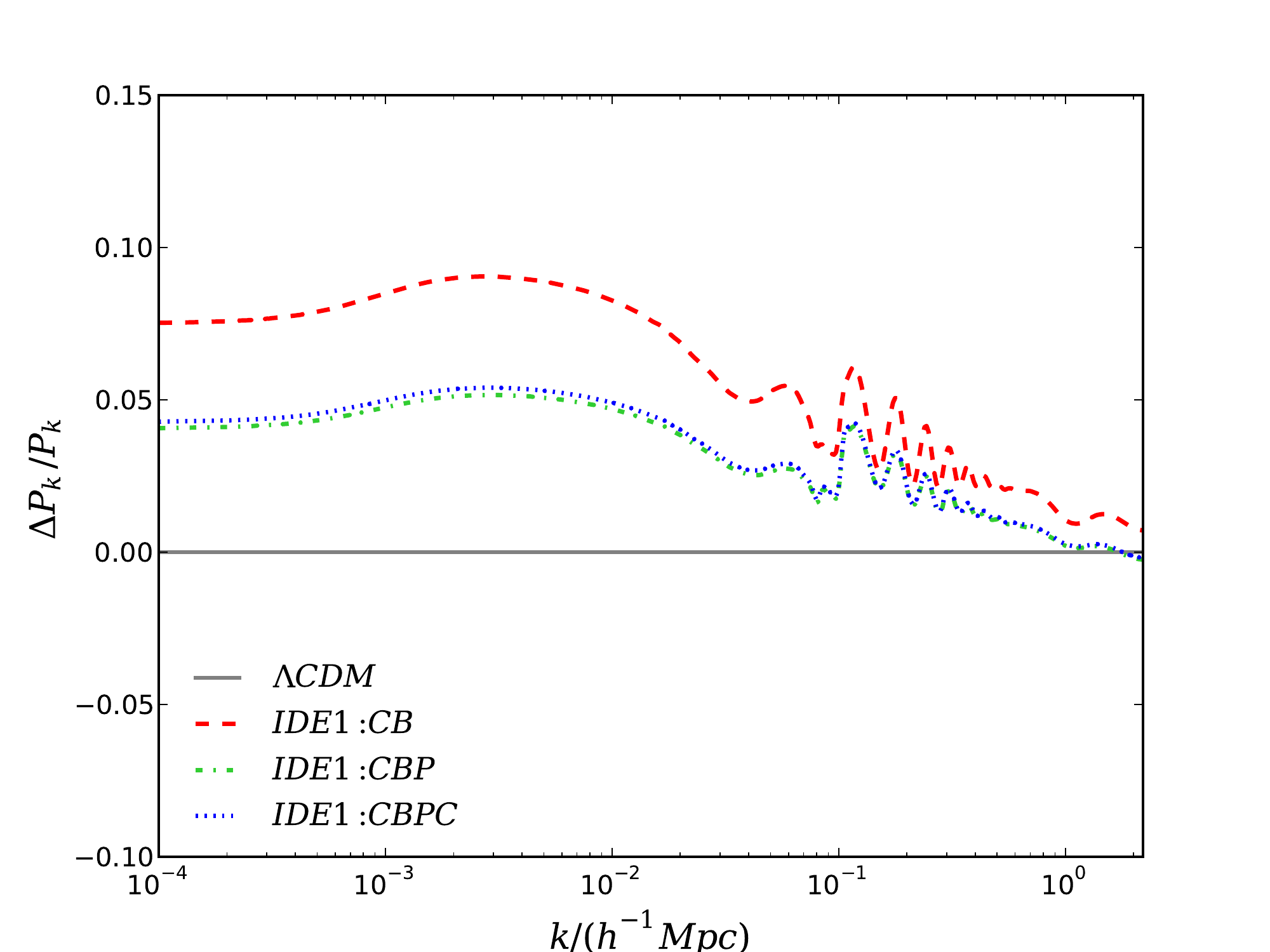}
\caption{{\it{
The   matter power spectra  for the interaction 
model IDE1 of (\ref{IDE1}), considering three different combinations of the observational 
datasets, namely CB ($=$ CMB+BAO), CBP ($=$CMB+BAO+Pantheon)  and CBPC ($=$ 
CMB+BAO+Pantheon+CC), as well as the   curve for $\Lambda$CDM paradigm (upper 
graph), and the corresponding residual plot with reference to 
$\Lambda$CDM scenario (lower graph). }} }
\label{fig-mpower-ide1}
\end{figure}

\begin{figure*}
\includegraphics[width=0.6\textwidth]{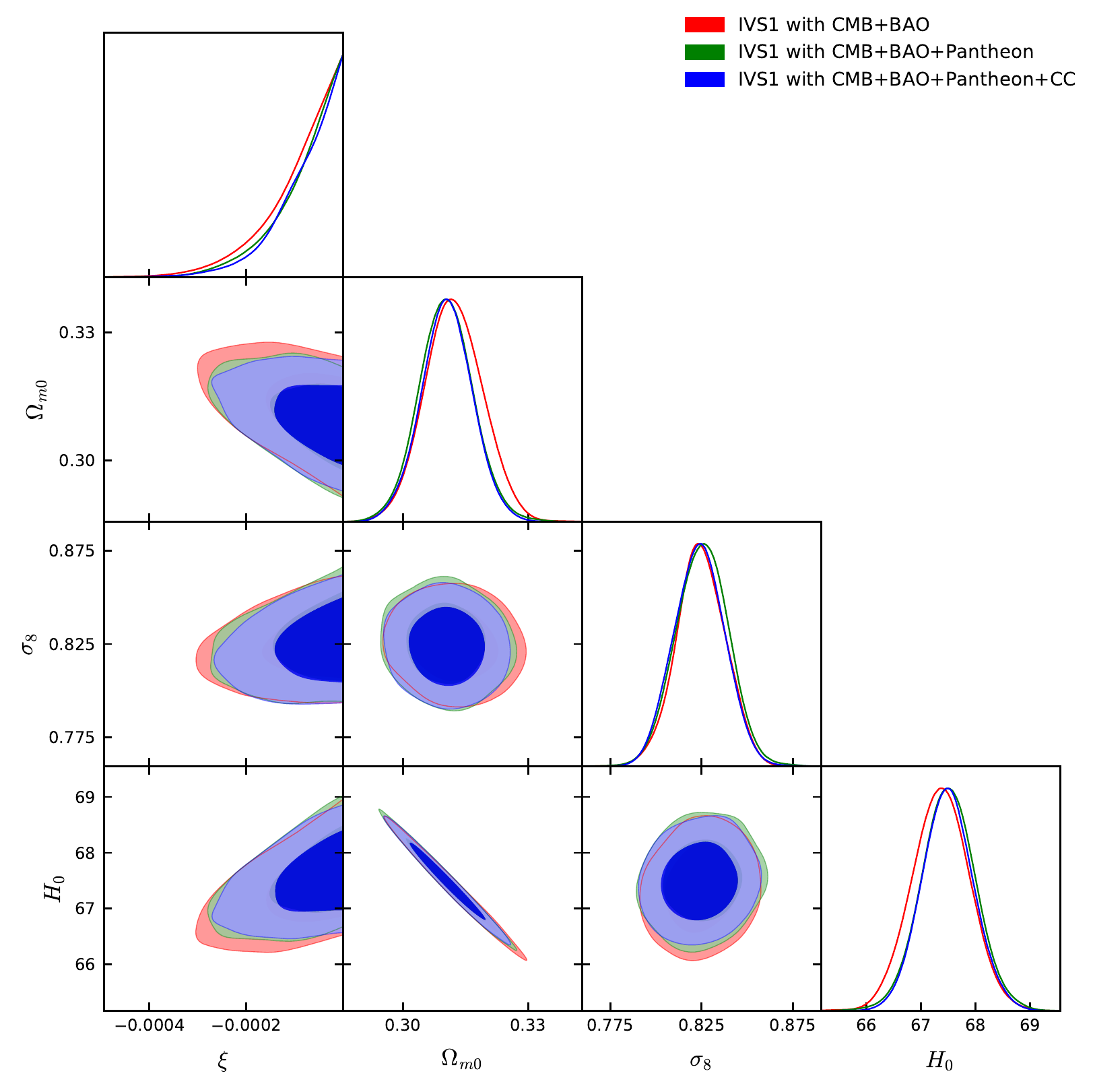}
\caption{{\it{The $1 \sigma$ and  $ 2 \sigma$ CL  
 contour 
plots for several combinations of various quantities and using
various combinations of the observational data sets, 
for the interaction model IVS1, and the corresponding 1D 
marginalized posterior distributions.
 Here, $\Omega_{m0}$ denotes the present value of $\Omega_m = 
\Omega_b+\Omega_c$ and $H_0$ is in the units of km/s/Mpc. }} }
\label{contour:ivs1}
\end{figure*}

\section{Observational Constraints}
\label{sec-results}

In this section we provide the observational constraints on the sign-changeable scenarios 
and we 
discuss their consequences.  In order to acquire a complete picture we consider three 
different 
combinations of the observational
datasets described above, namely CMB+BAO, CMB+BAO+Pantheon and CMB+BAO+Pantheon+CC.

\subsection{Interaction function  $Q = 3 H \xi (\rho_c - \rho_x)$}
\label{sec-results-model1}

The observational summary for IDE1 model is presented in Table 
\ref{tab:constraints-Model1}. Additionally, in   Fig. \ref{contour:ide1} we provide the 
corresponding 2D contour plots of various parameters at  1$\sigma$ and 2$\sigma$  
confidence level (CL), alongside the 1D marginalized posterior distribution.
From Table \ref{tab:constraints-Model1}   we can notice  that 
the 
addition of Pantheon or Pantheon+CC to the observational combination CMB+BAO, slightly 
improves the constraints by reducing their error bars, nevertheless the improvement is 
not   significant.  Additionally, from 
 Fig. \ref{contour:ide1} we can see that the parameters ($H_0$, $w_x$) and ($H_0$, 
$\Omega_{m0}$) are negatively correlated to each other. 

Concerning the dark energy equation-of-state parameter $w_x$, from 
Fig. \ref{contour:ide1} we infer that a phantom nature of $w_x$ is clearly preferred at 
more than 2$\sigma$ CL.  Additionally, concerning the coupling  $\xi$ and the dark 
energy equation-of-state parameter $w_x$, from Table  \ref{tab:constraints-Model1},  
it is clearly observed that the evidence for an interaction is  small ($\xi =  
-0.00014_{- 
   0.00004}^{+    0.00014}$ at 1$\sigma$ for CMB+BAO, $\xi =  
-0.00013_{-    0.00003}^{
+  0.00013}$ at 1$\sigma$ for  CMB+BAO+Pantheon, $\xi = -0.00013_{-    0.00003}^{+    
0.00013}$  at 1$\sigma$ for CMB+BAO+Pantheon+CC). Hence, within $1\sigma$ the 
non-interacting cosmology is   allowed.
In summary, our observational confrontation shows that this interaction 
model at the background level essentially mimicks a $w_x$CDM-type cosmology with $w_x < 
-1$ at more than 2$\sigma$ CL.

 We now proceed by examining the tensions on the two main parameters, namely $H_0$ and 
$\sigma_8$.  As we observe from both Table  \ref{tab:constraints-Model1} and  Fig. 
\ref{contour:ide1} (specifically from the posterior distribution of $H_0$ which is the 
extreme right plot of the bottom 
panel), $H_0$ acquires slightly higher values compared to  Planck \cite{Ade:2015xua}, 
with 
slightly higher error bars. Although the local estimation of the Hubble constant 
obtained by Riess et al. \cite{Riess:2016jrr}, i.e. $H_0 = 73.24\pm 1.74$, is certainly 
greater than the 
estimated mean values of $H_0$  for this interaction model, due to the increased error 
bars  the tension is reduced  to the level of $2\sigma$ CL.  Thus,   the interaction 
between the dark components  provides a way   to reduce the tension from 
$> 3 \sigma$ CL. Nevertheless, concerning  $\sigma_8$ tension
we deduce that the  present interaction model  is not able to alleviate it.

Let us now examine the effect  of the interaction on the large scale observables, and 
mainly on the CMB TT and matter power spectra. In the 
upper graph of Fig. \ref{fig-cmb-ide1} we show how the interaction affects the CMB TT 
spectra, considering the constraints on the parameters extracted from  all   
observational datasets, namely CMB+BAO, CMB+BAO+Pantheon and CMB+BAO+Pantheon+CC, in 
which for completeness  we add the non-interacting case of
$\Lambda$CDM cosmology. From this graph it is hard to distinguish the interacting case 
from $\Lambda$CDM  scenario. However, in the lower graph of Fig. 
\ref{fig-cmb-ide1} we depict the corresponding residual plot (with reference to 
$\Lambda$CDM model), and one can indeed trace a distinction between   interacting and 
non-interacting cosmologies, mainly  in the lower multipoles. 
Similarly, we investigate the effects of the interaction through the matter power spectra 
presented in Fig. \ref{fig-mpower-ide1}. Although in the upper graph the distinction 
between   interacting and non-interacting cosmologies cannot be observed, in the lower 
graph the deviation from the non-interacting $\Lambda$CDM cosmology is clear.  This is 
one 
of the main results of the
present work.

We close the analysis of this model by focusing on the case where $w_x = -1$, thus $w_x$ 
is not a free parameter and is fixed to the cosmological constant value,  namely we 
examine the interacting vacuum 
scenario. In this case we summarize the results in 
Table \ref{tab:cons-ivs1}, and in Fig. \ref{contour:ivs1}  we present the corresponding 
contour plots. In this model the case of no-interaction seems to be favored, while 
$H_0$ acquires smaller values, and therefore the $H_0$ tension is not alleviated. 
Concerning $\sigma_8$ tension we deduce that it  cannot be released either. 
Finally, at large scales this specific interacting scenario 
does not return different results in the CMB TT and matter power spectra comparing to  
$\Lambda$CDM cosmology, and therefore we do not explicitly present the corresponding 
plots.

\begin{center}             
\begin{table*}             
\begin{tabular}{cccccccccccccc}        
\hline\hline                         
Parameters & CMB+BAO & CMB+BAO+Pantheon & CMB+BAO+Pantheon+CC  \\ \hline
$\Omega_c h^2$ & $   0.1127_{-    0.0037-    0.0133}^{+    0.0080+    0.0099}$ & $    
0.1149_{-    
0.0025-    0.0078}^{+    0.0048+    0.0066}$ & $   0.1145_{-    0.0029-    0.0078}^{+    
0.0051+    
0.0069 }$ \\

$\Omega_b h^2$ & $    0.02229_{-    0.00015-    0.00029}^{+    0.00015+    0.00032}$ &  $ 
 
  0.
02229_{-    0.00015-    0.00031}^{+    0.00015+    0.00031}$ & $    0.02230_{-    
0.00014- 
   0.
00029}^{+    0.00014+    0.00031}$ \\

$100\theta_{MC}$ & $    1.04090_{-    0.00052-    0.00093}^{+    0.00042+    0.00097}$ & 
$ 
   1.
04081_{-    0.00036-    0.00070}^{+    0.00036+    0.00071}$ & $    1.04083_{-    
0.00040- 
   0.
00070}^{+    0.00036+    0.00077}$ \\

$\tau$ &  $    0.079_{-    0.017-    0.033}^{+    0.017+    0.032}$ & $    0.081_{-    
0.017-    0.
034}^{+    0.017+    0.032}$ & $    0.081_{-    0.017-    0.033}^{+    0.017+    0.032}$ 
\\

$n_s$ & $    0.9731_{-    0.0037-    0.0075}^{+    0.0037+    0.0073}$ &  $    0.9738_{-  
 
 0.0041- 
   0.0074}^{+    0.0038+    0.0080}$ & $    0.9736_{-    0.0044-    0.0079}^{+    0.0038+ 
 
  0.0081 
}$ \\

${\rm{ln}}(10^{10} A_s)$ & $    3.101_{-    0.032-    0.065}^{+    0.032+    0.062}$ & $  
 
 3.104_{-
    0.032-    0.067}^{+    0.033+    0.062}$ & $    3.105_{-    0.033-    0.066}^{+    
0.033+    0.
063}$ \\

$w_x$ & $   -1.0931_{-    0.0374-    0.0865}^{+    0.0529+    0.0777}$  & $   -1.0549_{-  
 
 0.0181- 
   0.0517}^{+    0.0320+    0.0455}$ & $   -1.0536_{-    0.0186-    0.0496}^{+    0.0290+ 
 
  0.0449}
$ \\

$\alpha$ & $   -0.00015_{-    0.00004-    0.00019}^{+    0.00015+    0.00015}$ & $   
-0.00014_{-    
0.00003-    0.00021}^{+    0.00014+    0.00014}$ & $   -0.00014_{-    0.00003-    
0.00020}^{+    0.
00014+    0.00014}$ \\

$\beta$ & $   -0.02444_{-    0.00681-    0.04197}^{+    0.02444+    0.02444}$ & $   
-0.01649_{-    
0.00394-    0.02375}^{+    0.01649+    0.01649}$ & $   -0.01783_{-    0.00529-    
0.02396}^{+    0.
01783+    0.01783}$ \\

$\Omega_{m0}$ &  $    0.279_{-    0.015-    0.039}^{+    0.023+    0.034}$ & $    
0.292_{- 
   0.010-
    0.024}^{+    0.013+    0.023}$ & $    0.291_{-    0.011-    0.026}^{+    0.014+    
0.023}$ \\

$\sigma_8$ &  $    0.906_{-    0.072-    0.093}^{+    0.032+    0.122}$ & $    0.875_{-   
 
0.040-   
 0.061}^{+    0.023+    0.069}$ & $    0.879_{-    0.043-    0.063}^{+    0.025+    
0.072}$ \\

$H_0$ & $   69.81_{-    1.39-    2.40}^{+    1.18+    2.51}$ & $   68.68_{-    0.73-    
1.41}^{+    
0.73+    1.41}$ & $   68.72_{-    0.80-    1.37}^{+    0.70+    1.47}$ \\
\hline\hline                
\end{tabular}               
\caption{Summary of the  $1 \sigma$ and  $ 2 \sigma$  CL 
constraints on  the interaction model IDE2 of (\ref{IDE2}),
using various combinations of the observational 
data sets. Here, $\Omega_{m0}$ denotes the present value of $\Omega_m = 
\Omega_b+\Omega_c$ and $H_0$ is in the units of km/s/Mpc.
}
\label{tab:constraints-Model2}   
\end{table*}                 
\end{center}

\begin{figure*}
\includegraphics[width=0.6\textwidth]{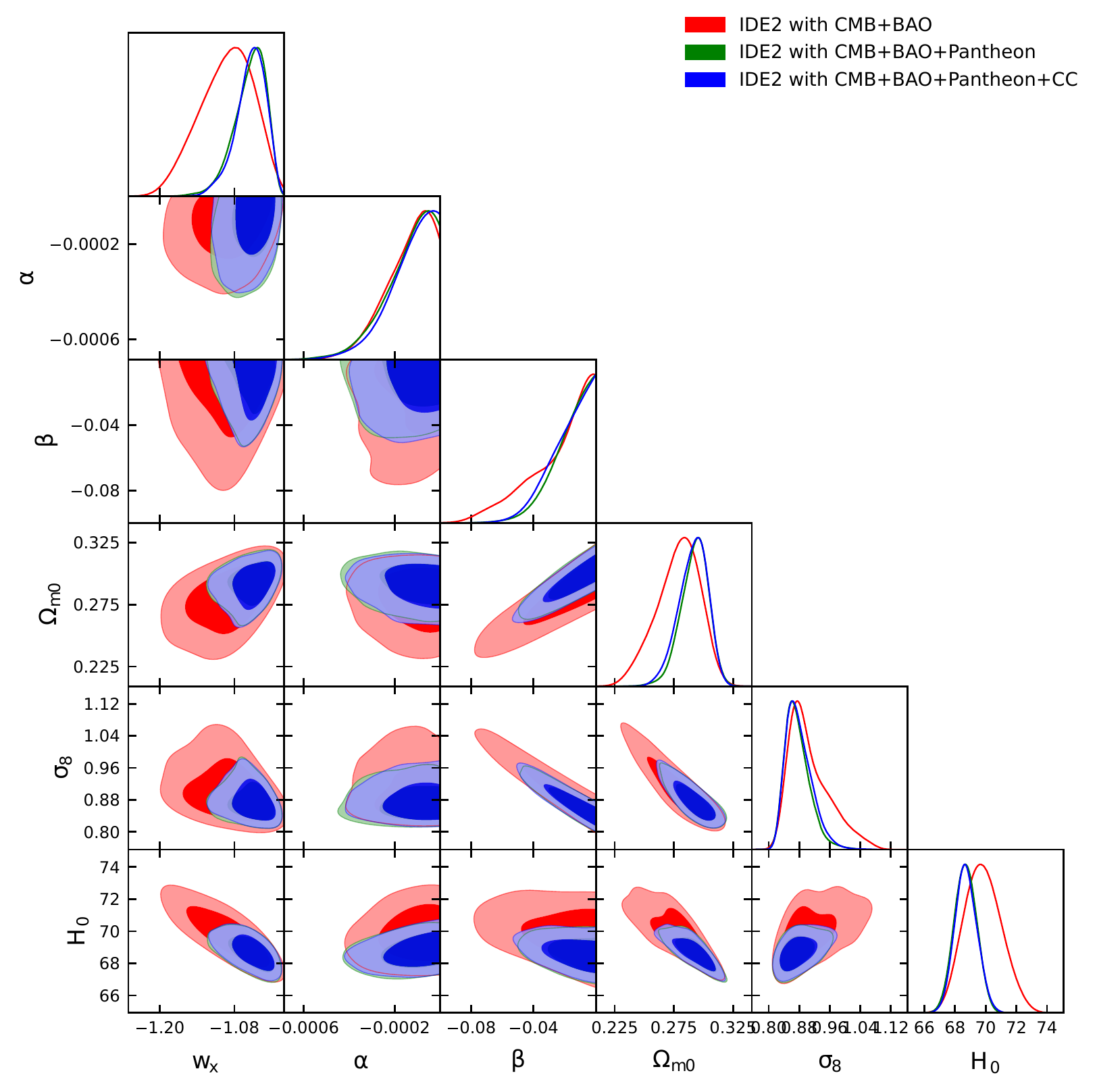}
\caption{{\it{The $1 \sigma$ and  $ 2 \sigma$ CL  
  contour 
plots for several combinations of various quantities and using
various combinations of the observational data sets, 
for the interaction model IDE2 (\ref{IDE2}), and the corresponding 1D 
marginalized posterior distributions.
 Here, $\Omega_{m0}$ denotes the present value of $\Omega_m = 
\Omega_b+\Omega_c$ and $H_0$ is in the units of km/s/Mpc.}} }
\label{contour:ide2}
\end{figure*}

\begin{figure}[ht]
\includegraphics[width=0.4\textwidth]{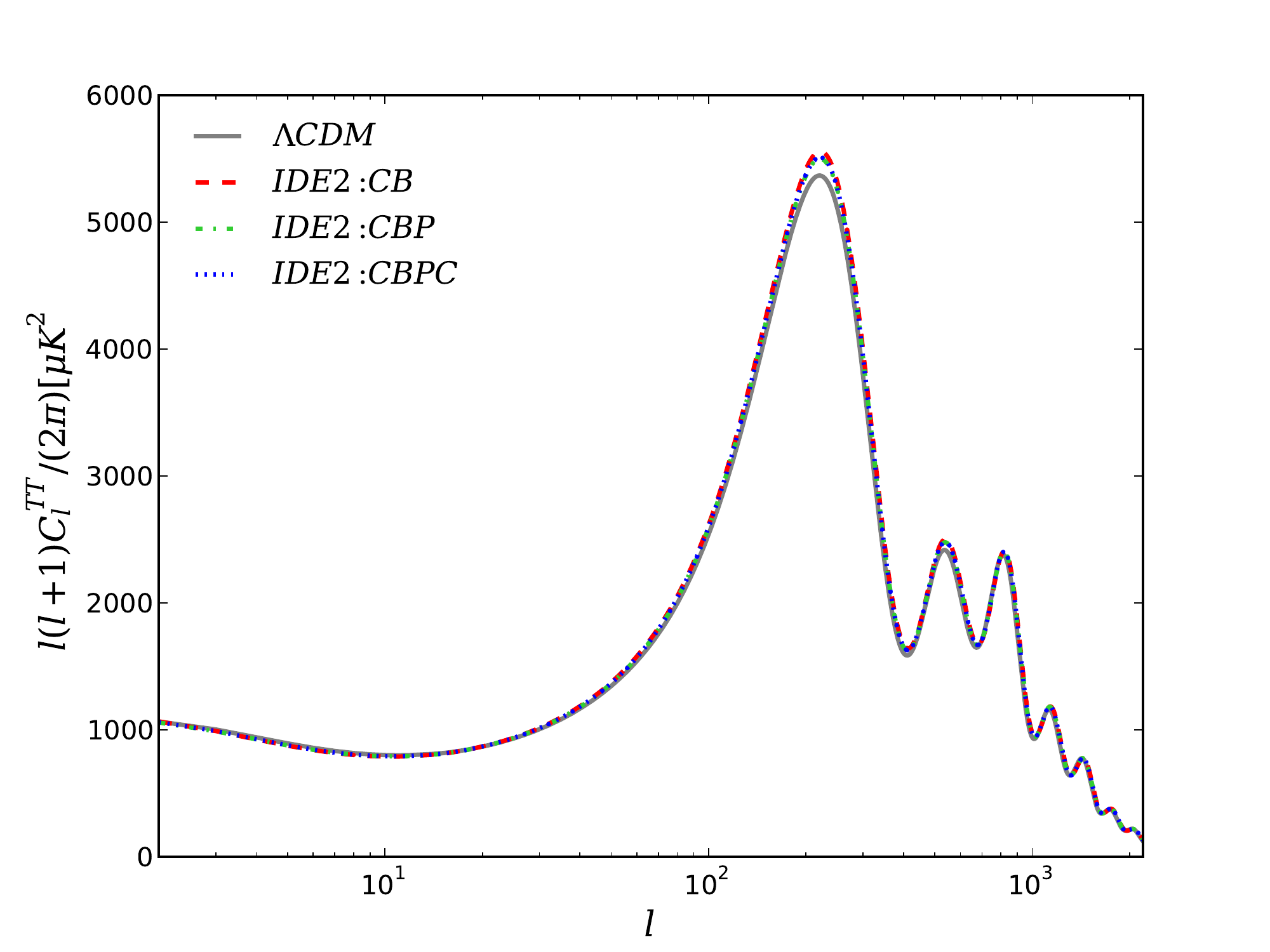}
\includegraphics[width=0.4\textwidth]{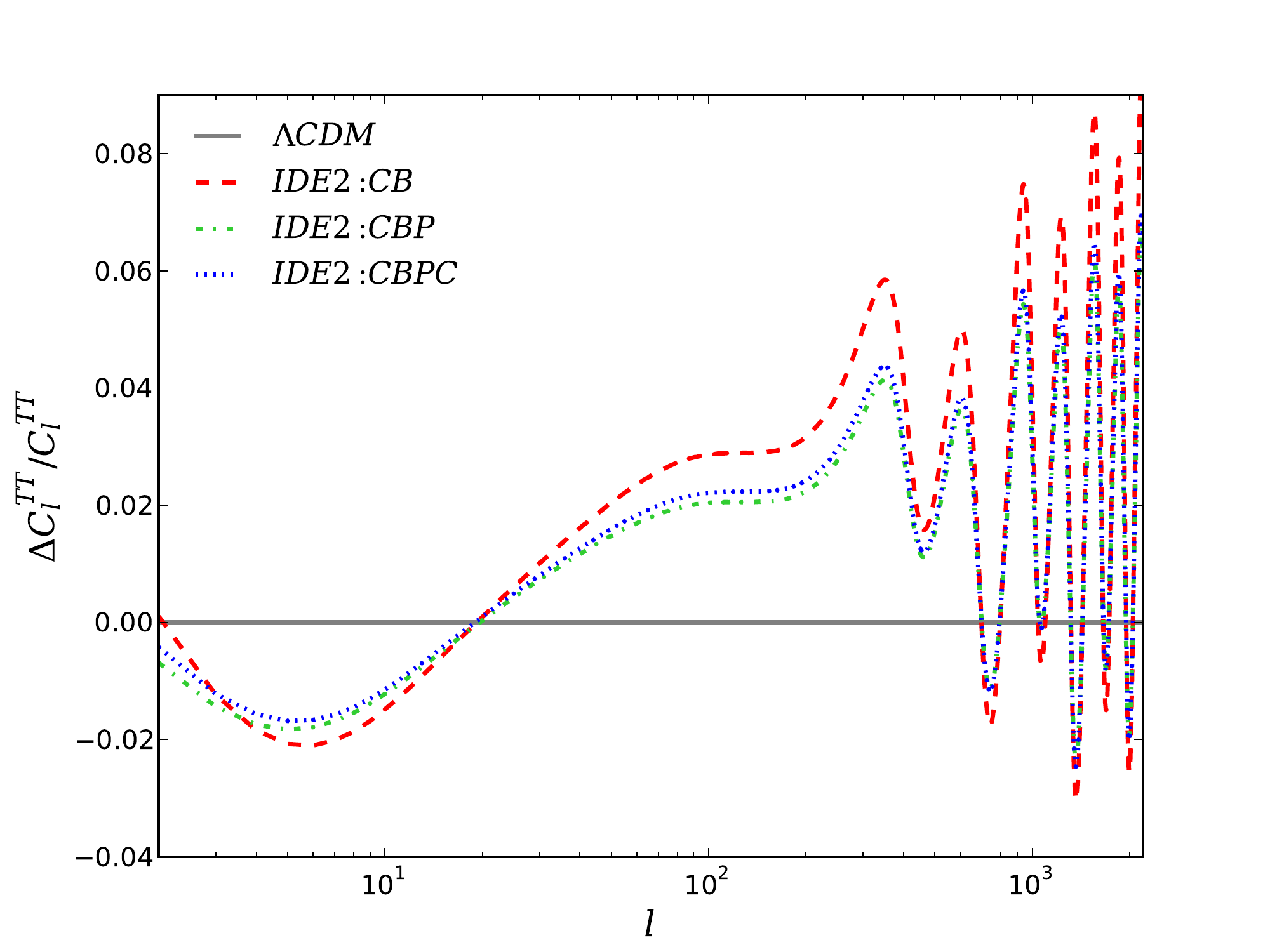}
\caption{{\it{
The temperature anisotropy in the CMB TT spectra for the interaction 
model IDE2 of (\ref{IDE2}), considering three different combinations of the observational 
datasets, namely CB ($=$ CMB+BAO), CBP ($=$CMB+BAO+Pantheon)  and CBPC ($=$ 
CMB+BAO+Pantheon+CC), as well as the   curve for $\Lambda$CDM paradigm (upper 
graph), and the corresponding residual plot with reference to 
$\Lambda$CDM scenario (lower graph).}} }
\label{fig-cmb-ide2}
\end{figure}

\begin{figure}[ht]
\includegraphics[width=0.4\textwidth]{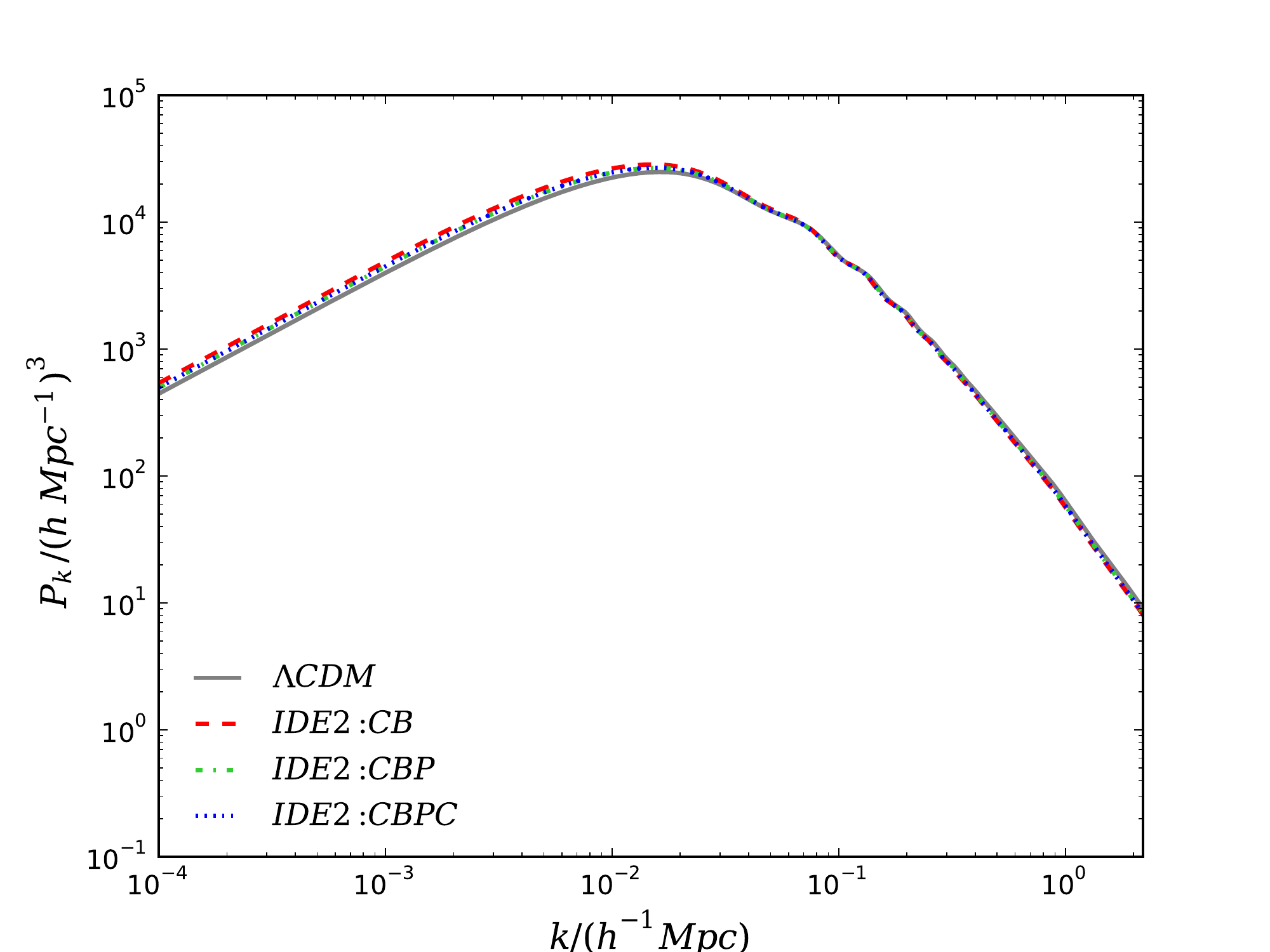}
\includegraphics[width=0.4\textwidth]{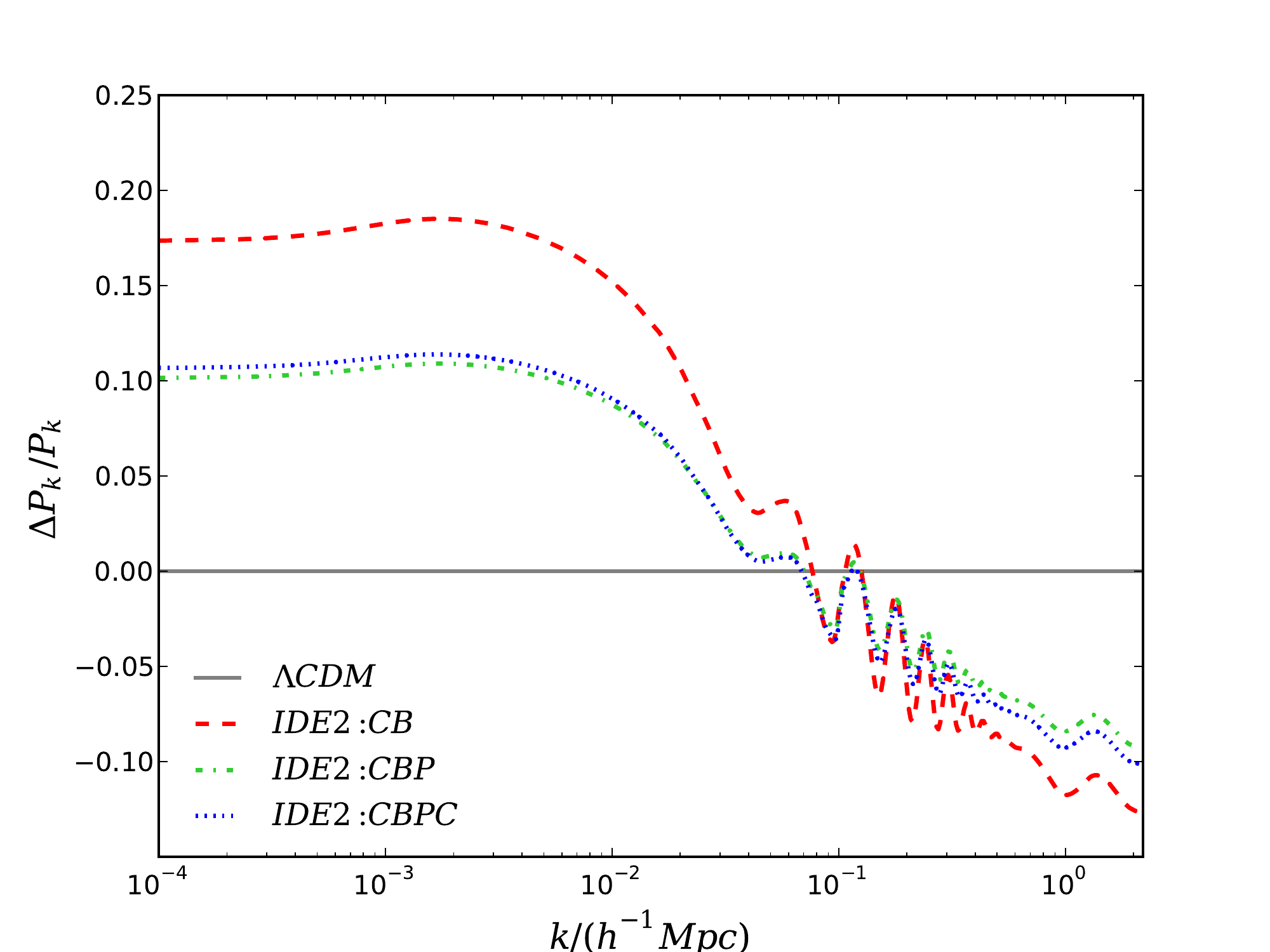}
\caption{{\it{
The   matter power spectra  for the interaction 
model IDE2 of (\ref{IDE2}), considering three different combinations of the observational 
datasets, namely CB ($=$ CMB+BAO), CBP ($=$CMB+BAO+Pantheon)  and CBPC ($=$ 
CMB+BAO+Pantheon+CC), as well as the   curve for $\Lambda$CDM paradigm (upper 
graph), and the corresponding residual plot with reference to 
$\Lambda$CDM scenario (lower graph). }} }
\label{fig-mpower-ide2}
\end{figure}

\subsection{Interaction function  $Q = 3 H (\alpha \rho_c - \beta \rho_x)$}
\label{sec-results-model2}

The observational summary  of   model IDE2 is shown in Table \ref{tab:constraints-Model2},
while the corresponding 2D contour plots are presented in Fig. \ref{contour:ide2}. 
Similarly to model IDE1,  one can notice that the addition of Pantheon or Pantheon+CC to 
the combined analysis CMB+BAO, improves the parameters space only slightly. Moreover, 
from 
 Fig. \ref{contour:ide2} we can see that the parameters ($H_0$, $w_x$) and ($H_0$, 
$\Omega_{m0}$) are negatively correlated to each other.

 Concerning the coupling parameters    $\alpha$ and $\beta$, our analysis shows 
that  the 
zero values are allowed within 1$\sigma$. Additionally, the dark-energy 
equation-of-state parameter prefers the phantom regime 
for all   datasets, namely we see that $w_x < -1$ at more 
than 2$\sigma$. Furthermore, similarly to IDE1 model, in the present IDE2 scenario  we 
also find that the estimations of $H_0$ are slightly higher compared to the 
$\Lambda$CDM-based 
Planck estimation \cite{Ade:2015xua}, and due to the higher error bars on $H_0$ the 
relevant tension can be slightly reconciled due to the interaction.
However, concerning $\sigma_8$ we see that the tension is not released.

We proceed by  investigating the effect of the interaction  on the CMB TT and matter 
power 
spectra. In the upper graph of Fig. \ref{fig-cmb-ide2} we depict the CMB TT spectra 
considering the constraints on the parameters extracted from  all observational 
datasets, namely CMB+BAO, CMB+BAO+Pantheon, and CMB+BAO+Pantheon+CC, in which for 
completeness  we add the non-interacting case of $\Lambda$CDM cosmology. Moreover, in the 
lower graph of Fig. \ref{fig-cmb-ide2} we present the corresponding residual plot (with 
reference to $\Lambda$CDM model).
As we observe,  this interaction model is distinguished from the non-interacting 
$\Lambda$CDM cosmology at both lower and higher multipoles. 
We further investigate the effects of the interaction on the matter 
power spectra, depicted in Fig. \ref{fig-mpower-ide2}. Although in the upper graph the 
distinction between the interacting and non-interacting cosmologies cannot be observed, 
in 
the lower graph the deviation from the non-interacting $\Lambda$CDM cosmology is clear
even for the small values of the coupling parameters $\alpha$, $\beta$ that were obtained 
from the  three different observational datasets. This is one of the main results of the 
present work.

\begin{center}                    
\begin{table*}               
\begin{tabular}{cccccccccccccc}        
\hline\hline                  
Parameters & CMB+BAO & CMB+BAO+Pantheon & CMB+BAO+Pantheon+CC  \\ \hline
$\Omega_c h^2$ & $    0.1141_{-    0.0020-    0.0085}^{+    0.0048+    0.0069}$ & $    
0.1085_{-    
0.0044-    0.0159}^{+    0.0101+    0.0126}$  & $    0.1055_{-    0.0053-    0.0181}^{+   
 
0.0111+  
  0.0149}$ \\
$\Omega_b h^2$ & $    0.02234_{-    0.00013-    0.00028}^{+    0.00013+    0.00026}$ & $  
 
 0.02233_
{-    0.00015-    0.00030}^{+    0.00016+    0.00030}$  & $    0.02231_{-    0.00015-    
0.00029}^{
+    0.00014+    0.00030}$\\
$100\theta_{MC}$ & $    1.04088_{-    0.00037-    0.00077}^{+    0.00035+    0.00086}$ & 
$ 
   1.
04119_{-    0.00058-    0.00102}^{+    0.00044+    0.00107}$ & $    1.04135_{-    
0.00069- 
   0.
00109}^{+    0.00045+    0.00123}$\\
$\tau$ & $    0.087_{-    0.016-    0.035}^{+    0.018+    0.033}$ & $    0.085_{-    
0.017-    0.
032}^{+    0.017+    0.032}$ & $    0.084_{-    0.017-    0.034}^{+    0.017+    0.032}$\\
$n_s$ & $    0.9766_{-    0.0037 -.0080}^{+    0.0036+    0.0079}$ & $    0.9758_{-    
0.0039-    0.
0076}^{+    0.0040+    0.0079}$ & $    0.9753_{-    0.0038-    0.0076}^{+    0.0038+    
0.0076}$ \\
${\rm{ln}}(10^{10} A_s)$ & $    3.114_{-    0.031-    0.069}^{+    0.036+    0.062}$ & $  
 
 3.111_{-
    0.032-    0.062}^{+    0.032+    0.064}$ & $    3.109_{-    0.033-    0.066}^{+    
0.034+    0.
064}$\\
$\alpha$ & $   -0.000092_{-    0.000019-    0.000153}^{+    0.000092+    0.000092}$ & $   
-0.000099_
{-    0.000023-    0.000157}^{+    0.000099+    0.000099}$ & $   -0.000094_{-    
0.000022- 
   0.
000147}^{+    0.000094+    0.000094}$ \\

$\beta$ & $   -0.01474_{-    0.00223-    0.02488}^{+    0.01474+    0.01474}$ & $   
-0.03330_{-    
0.00844-    0.04914}^{+    0.03330+    0.03330}$ & $   -0.04238_{-    0.01469-    
0.05246}^{+    0.
03732+    0.04238}$\\

$\Omega_{m0}$ & $    0.299_{-    0.009-    0.027}^{+    0.014+    0.025}$ & $    0.284_{- 
 
  0.015- 
   0.042}^{+    0.026+    0.036}$ & $    0.276_{-    0.016-    0.048}^{+    0.029+    
0.042}$ \\
$\sigma_8$ & $    0.857_{-    0.037-    0.058}^{+    0.019+    0.068}$  & $    0.902_{-   
 
0.084-   
 0.106}^{+    0.032+    0.143}$ & $    0.927_{-    0.093-    0.127}^{+    0.039+    
0.163}$\\
$H_0$ & $   67.71_{-    0.57-    1.15}^{+    0.61+    1.11}$ & $   68.05_{-    0.64-    
1.29}^{+    
0.67+    1.24}$  & $   68.24_{-    0.80-    1.30}^{+    0.64+    1.50}$\\
\hline\hline               
\end{tabular}               
\caption{Summary of the  $1 \sigma$ and  $ 2 \sigma$  CL 
constraints on  the interaction model IVS2,
using various combinations of the observational 
data sets. Here, $\Omega_{m0}$ denotes the present value of $\Omega_m = 
\Omega_b+\Omega_c$ and $H_0$ is in the units of km/s/Mpc.
}\label{tab:cons-ivs2}                                     
\end{table*}                       
\end{center}

\begin{figure*}
\includegraphics[width=0.6\textwidth]{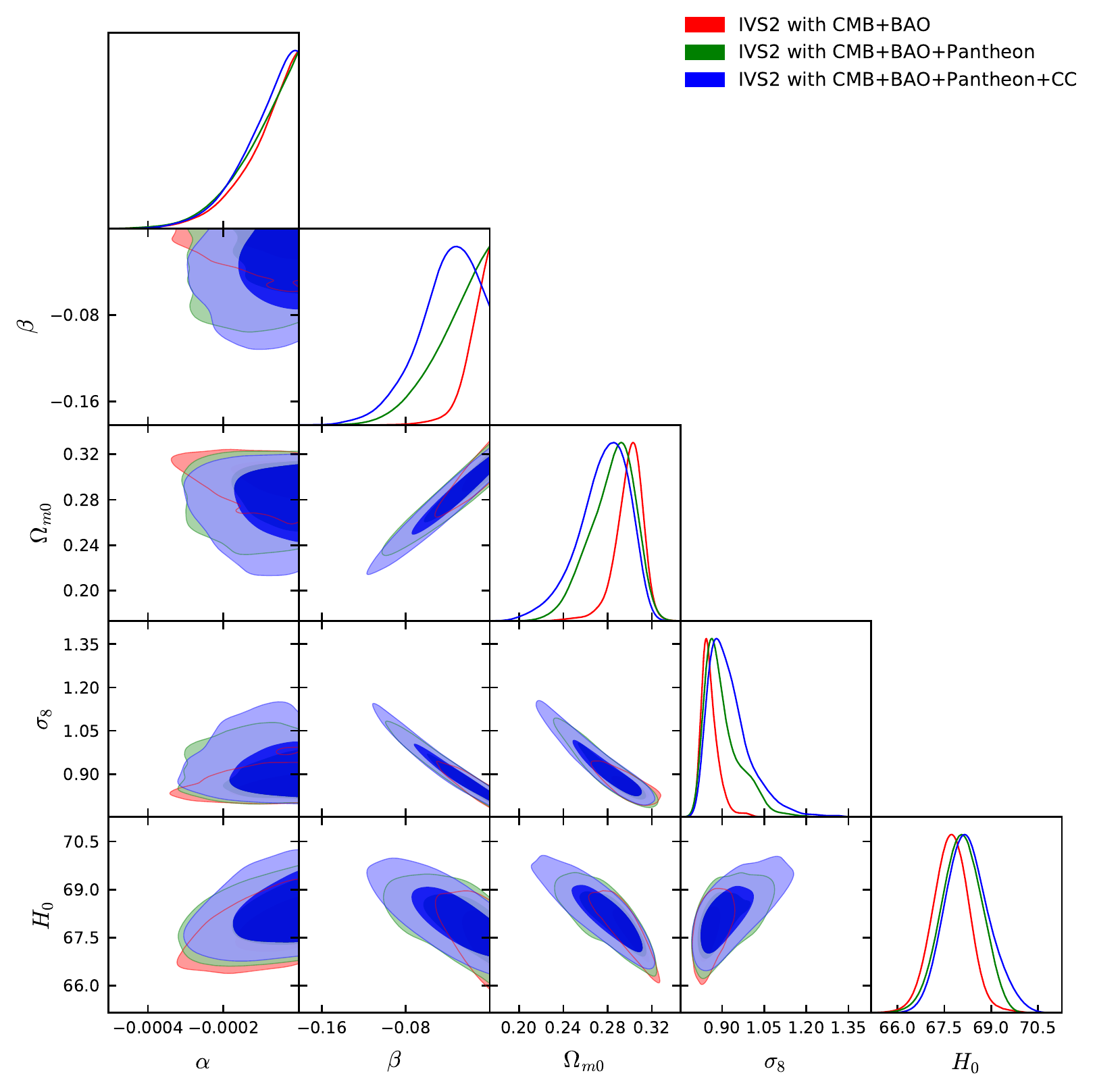}
\caption{{\it{
The $1 \sigma$ and  $ 2 \sigma$ CL  
  contour 
plots for several combinations of various quantities and using
various combinations of the observational data sets, 
for the interaction model IVS2, and the corresponding 1D 
marginalized posterior distributions.
 Here, $\Omega_{m0}$ denotes the present value of $\Omega_m = 
\Omega_b+\Omega_c$ and $H_0$ is in the units of km/s/Mpc.}} }
\label{contour:ivs2}
\end{figure*}
 
We proceed by analyzing the case $w_x = -1$, namely the interacting vacuum scenario. The 
results of the analyses are presented in Table \ref{tab:cons-ivs2}, while in Fig. 
\ref{contour:ivs2} we depict the corresponding contour plots. From these we deduce that 
the 
coupling parameter $\alpha$ acquires the zero value 
within $1\sigma$ CL irrespectively of 
the datasets, while $\beta$ has a tendency towards non-zero values nevertheless the 
value zero is allowed within 1$\sigma$ except for the final combination 
CMB+BAO+Pantheon+CC. Additionally, concerning  $H_0$, for 
CMB+BAO+Pantheon and CMB+BAO+Pantheon+CC datasets we find that its estimations are 
relatively high compared to  $\Lambda$CDM-based Planck estimation \cite{Ade:2015xua}, 
while the error bars on $H_0$ are also increased. Hence, this enables $H_0$ to acquire  
values close to its local estimation \cite{Riess:2016jrr}, and thus the tension on $H_0$ 
is weakly resolved. However, the $\sigma_8$ tension  cannot be alleviated.

\subsection{Bayesian analysis}
\label{sec-bayesian} 

We close the observational confrontation by   presenting   observational viabilities 
of the models using the Bayesian evidence. The Bayesian analysis is an important part of 
the cosmological model selection that quantifies the fitting results  
compared to a reference scenario. The 
computation of Bayesian evidence is  performed with the code \texttt{MCEvidence} 
\cite{Heavens:2017hkr,Heavens:2017afc}\footnote {See the freely available code in
\href{https://github.com/yabebalFantay
e/MCEvidence}{github.com/yabebalFantaye/MCEvidence}.}, which directly computes the 
evidences of the model   with respect to the reference  
$\Lambda$CDM scenario.  

In Bayesian analysis one needs to calculate the posterior probability of the model 
parameters $\theta$, subject to a particular observational dataset $x$ and any prior 
information for 
the underlying model $M$. Recalling the Bayes theorem one can write that  
\begin{eqnarray}\label{BE}
p(\theta|x, M) = \frac{p(x|\theta, M)\,\pi(\theta|M)}{p(x|M)},
\end{eqnarray}
in which  $p(x|\theta, M)$ is the likelihood function (depending on the 
model parameters $\theta$ with the given data set), and where $\pi(\theta|M)$ refers to 
the prior information. The quantity $p(x|M)$ is  the 
Bayesian evidence. Given two models, namely $M_i$ and $M_j$, where 
$M_i$ is the model under investigation and $M_j$ is the reference model (here the 
$\Lambda$CDM scenario), the posterior probability is given by 
\begin{eqnarray}
\frac{p(M_i|x)}{p(M_j|x)} = \frac{\pi(M_i)}{\pi(M_j)}\,\frac{p(x| M_i)}{p(x|M_j)} = 
\frac{\pi(M_i)}{
\pi(M_j)}\, B_{ij}. 
\end{eqnarray}
The quantity $B_{ij} = \frac{p(x| M_i)}{p(x|M_j)}$ is the Bayes factor of the considered 
model $M_i$ relative to the reference model $M_j$, and quantifies 
how the observational data support the model $M_i$ over $M_j$.  In 
Table \ref{tab:jeffreys} we show the corresponding classification following  
\cite{Kass:1995loi}.  
 
\begin{table}  [ht]                        
\begin{tabular}{ccc}      
\hline\hline                
$\ln B_{ij}$ & ~~~~~~~Strength of evidence for model ${M}_i$ \\ \hline
$0 \leq \ln B_{ij} < 1$ & Weak \\
$1 \leq \ln B_{ij} < 3$ & Definite/Positive \\
$3 \leq \ln B_{ij} < 5$ & Strong \\
$\ln B_{ij} \geq 5$ & Very strong \\
\hline\hline              
\end{tabular}     
\caption{The Revised Jeffreys scale from \cite{Kass:1995loi} that quantifies the fitting 
efficiency of the investigated model 
$M_i$ comparing to the reference model $M_j$.} 
\label{tab:jeffreys}     
\end{table}

\begin{table}      
\begin{center}                    
\begin{tabular}{cccccccc}                                      \hline\hline              
 
Dataset & Model & $\ \ \ln B_{ij}$  \\ 
\hline
CB & IDE1 & $-4.8$  \\
CBP & IDE1 & $-2.4$ \\
CBPC & IDE1 & $-2.9$ \\

\hline 

CB & IVS1 & $-3.9$  \\
CBP & IVS1 & $-2.9$  \\
CBPC & IVS1 & $-3.1$  \\ 

\hline
\hline 

CB & IDE2 & $-6.6$  \\
CBP & IDE2 & $-3.1$  \\
CBPC & IDE2 & $-4.2$  \\ 

\hline 

CB & IVS2 & $-5.1$  \\
CBP & IVS2 & $-3.6$  \\
CBPC & IVS2 & $-4.0$  \\ 

\hline\hline 
\end{tabular}    
\caption{The values of $\ln B_{ij}$ for all interaction scenarios comparing to the  
reference paradigm of $\Lambda$CDM, for all observational datasets. Here, CB $=$ CMB+BAO, 
CBP $=$ CMB+BAO+Pantheon and 
CBPC $=$ CMB+BAO+Pantheon+CC. 
The negative values of $\ln B_{ij}$ imply that the $\Lambda$CDM paradigm is 
preferred over the interaction scenarios. } 
\label{tab:bayesian}                          
\end{center}    
\end{table}  

In Table \ref{tab:bayesian} we present the computed values of $\ln B_{ij}$ for all   
sign-changeable interacting scenarios considering all   observational datasets. As we 
can see, the values of $|\ln B_{ij}|$ for IDE2 are greater than the values of   $|\ln 
B_{ij}|$ for IDE1 and this is true for all   datasets. This was expected since
IDE2 scenario has one extra free parameter compared to IDE1. Similarly,  the values of 
$|\ln B_{ij}|$ for model IVS2 are greater than those IVS1. Nevertheless, overall,  
$\Lambda$CDM cosmology   is still favored over the present interacting models.

\section{Laws of thermodynamics in sign-changeable interaction models}
\label{sec-thermo}

In this section we shall investigate the thermodynamical laws in a universe governed 
by  sign-changeable interacting dark energy.
In order to investigate the thermodynamical properties of a specific 
cosmological model one assumes that the universe is a thermodynamical system bounded by 
a cosmological horizon, and then he applies arguments from  black hole 
thermodynamics \cite{gibbons, jacobson, paddy}. In particular, 
one considers that the universe is bounded 
by the apparent horizon with radius $r_h= \left(H^2+ k/a^2  \right)^{-1/2}$ 
\cite{bak-rey} which, for a spatially flat universe  becomes the Hubble horizon $r_h = 
1/H$. Hence, one can show that the 
first law of thermodynamics  can lead to the first Friedmann equation  
\cite{horizon-temperature-4}. 

We proceed by examining the validity of the generalized second law of thermodynamics, 
which states that the total entropy of the universe, namely the entropy of the various 
cosmological fluids plus the entropy of the horizon, should be a non-decreasing function 
of time \cite{Bekenstein:1974ax}. 
Concerning the entropy of the various fluids that constitute the universe one has
$ S_r+ S_b 
+S_c+ S_x$,
 where $S_i$ ($i = r, b, c, x$) denotes the entropy of the $i$-th fluid. Thus, the
first law of thermodynamics for each individual fluid becomes
\cite{jacobson, paddy, bak-rey, horizon-temperature-1, 
horizon-temperature-2,  
horizon-temperature-3,  horizon-temperature-4, horizon-temperature-5,Jamil:2010di}.
\begin{align}
T dS_r &= dE_r + p_r dV ,\label{sp-thermo4}\\
T d S_b &= dE_b + p_b dV = dE_b,\label{sp-thermo5}\\
T dS_c &= dE_c + p_c dV =  dE_c,\label{sp-thermo6}\\
T d S_x &= dE_x + p_x dV,\label{sp-thermo7}
\end{align}
where $V= 4\pi r_h^3/ 3$ is the volume of the universe, $E_i $ 
stands for the internal energy of the $i$-th fluid given by $E_i = \frac{4}{3}\, \pi\, 
r_h^3\, \rho_i$, and $p_i$ is the corresponding pressure. Note that the various fluids 
are considered to have the same temperature. However, we mention that  
over the entire cosmic evolution the temperatures of different cosmic fluids 
are not the same \cite{Mimoso:2016jwg},  as they evolve differently. In 
particular, the temperatures of radiation and dark energy remain different from 
the horizon temperature for a long period of time, while for the 
non-relativistic matter its temperature becomes and remains equal with the 
horizon one. Nevertheless, if at some epoch in the universe evolution the 
horizon temperature comes close or become equal to  
the temperature of the dark energy sector, then they will remain the same for 
most of the expansion of the universe. Therefore, the assumption, that the 
temperatures of different cosmic fluids are the same, is not unjustified 
\cite{Mimoso:2016jwg}.

The entropy of the horizon is  
taken to be that of a black hole, namely
$
S_h= k_B \mathcal{A}/(4\; l_{pl}^2)
$,
with $k_B$   the Boltzmann's constant, $\hbar$  the Planck's constant, and $l_{pl} = 
(\sqrt{\hbar G/c^3})$  the Planck's length, however taking the area to be that 
corresponding to the horizon, i.e.  $\mathcal{A}= 4 \pi r_h^2$ \cite{gibbons, jacobson, 
paddy}. Therefore, in units where $\hbar= k_B= c= 8 \pi G = 1$, the horizon entropy 
(\ref{sp-thermo2}) reduces to
\begin{eqnarray}\label{sp-thermo2}
S_h = 8 \pi^2 r_h^2. 
\end{eqnarray}
Moreover, concerning the temperature of the apparent horizon one can use the black hole result but 
with the apparent horizon instead of the black-hole one, namely $T_h = 
1/(2 \pi r_h)$  \cite{jacobson, paddy, bak-rey, horizon-temperature-1, 
horizon-temperature-2,  
horizon-temperature-3,  horizon-temperature-4,Bekenstein:1974ax, 
horizon-temperature-5,Jamil:2010di}.

Using all the above relations together with the assumption that the temperature of 
the fluids  $T$ should be equal to $T_h$  one can 
find that 
\begin{eqnarray}\label{gslt}
&&\!\!\!\!\!\!
\dot{S}= \dot{S}_r+ \dot{S}_b+ \dot{S}_c + \dot{S}_x + \dot{S}_h \nonumber\\ 
&&= 4 \pi^2 H r_h^6 \Bigl[\rho_r (1+w_r) +\rho_b + \rho_c + (1+ w_x)\rho_x  \Bigr]^2,\ \ 
\ \ \,
\end{eqnarray}
where  we have replaced the involved $\dot{r}_h=-\dot{H}/H^2$ using the Friedmann 
equations (\ref{f1}),(\ref{f2}). As we can see, the total entropy is always a
non-decreasing function, and thus the generalized second law of thermodynamics is 
satisfied. Hence, although the sign change of the interaction between dark matter and dark 
energy could lead to a local entropy decrease at the microscopic level, the total entropy 
of the universe plus the one of the horizon is always non-decreasing.
The above relation for the first law of thermodynamics in interacting cosmology  has been 
established earlier in  \cite{Jamil:2009eb}.

Now, let us examine whether the universe with sign-changeable interaction in the dark 
sectors will result to thermodynamic equilibrium. As it is discussed extensively in the literature 
(see for instance \cite{Radicella:2010zb,delCampo:2012ya}), in order for this to be 
achieved one needs to have a total entropy whose first derivative is positive, while at 
late times its second derivative should be negative \cite{Gonzalez-Espinoza:2019vcy}, and thus the entropy asymptotically 
tends towards a constant value. Using $x= \ln a$ as the independent variable the above 
require that
$S^{\prime} \geq 
0$ and $S^{\prime \prime} < 0$, where $S=S_r+ S_b 
+S_c+ S_x+S_h$, with
primes denoting 
differentiation with respect to $x$.

\begin{figure}[ht]
\includegraphics[width=0.42\textwidth]{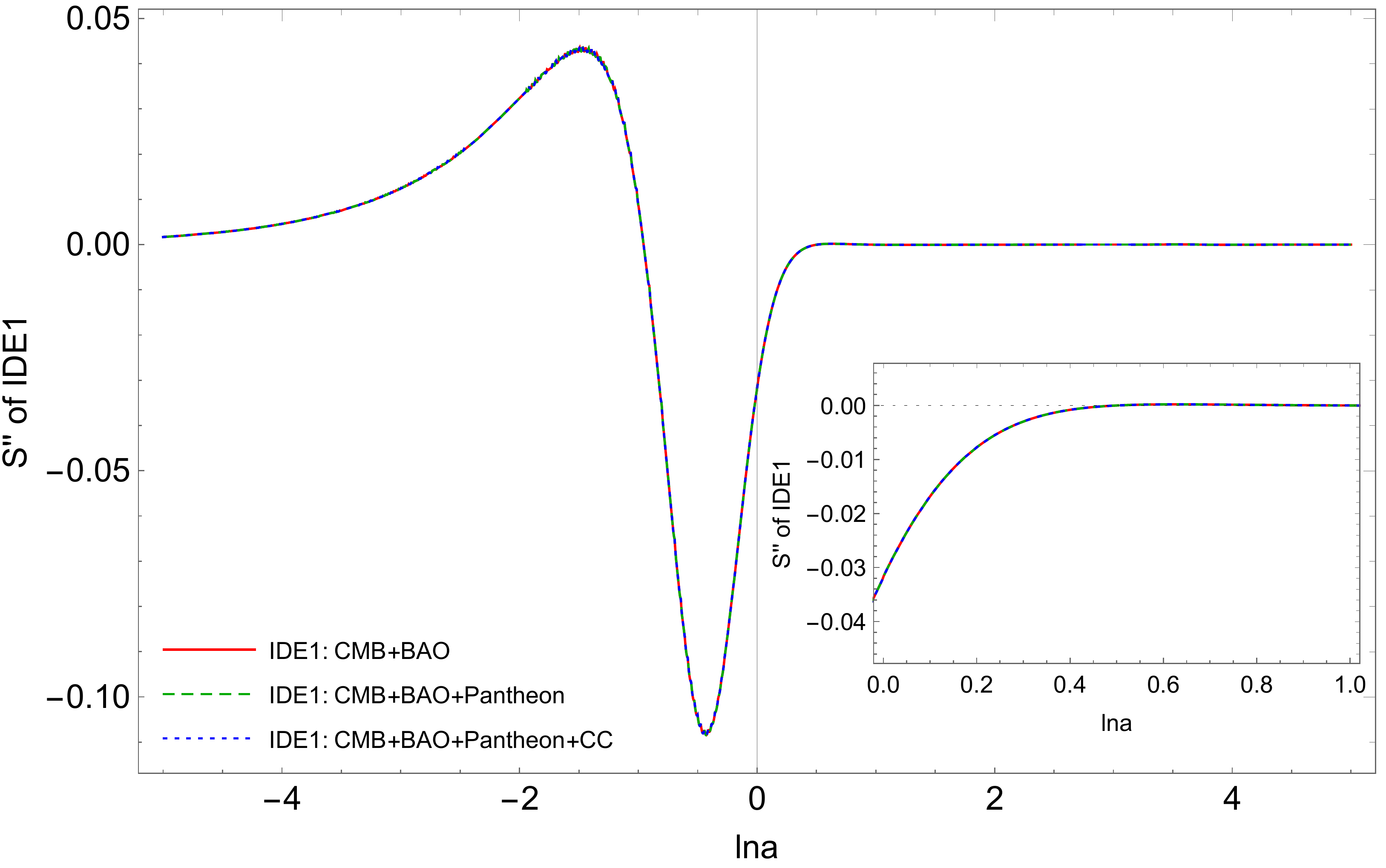}
\includegraphics[width=0.42\textwidth]{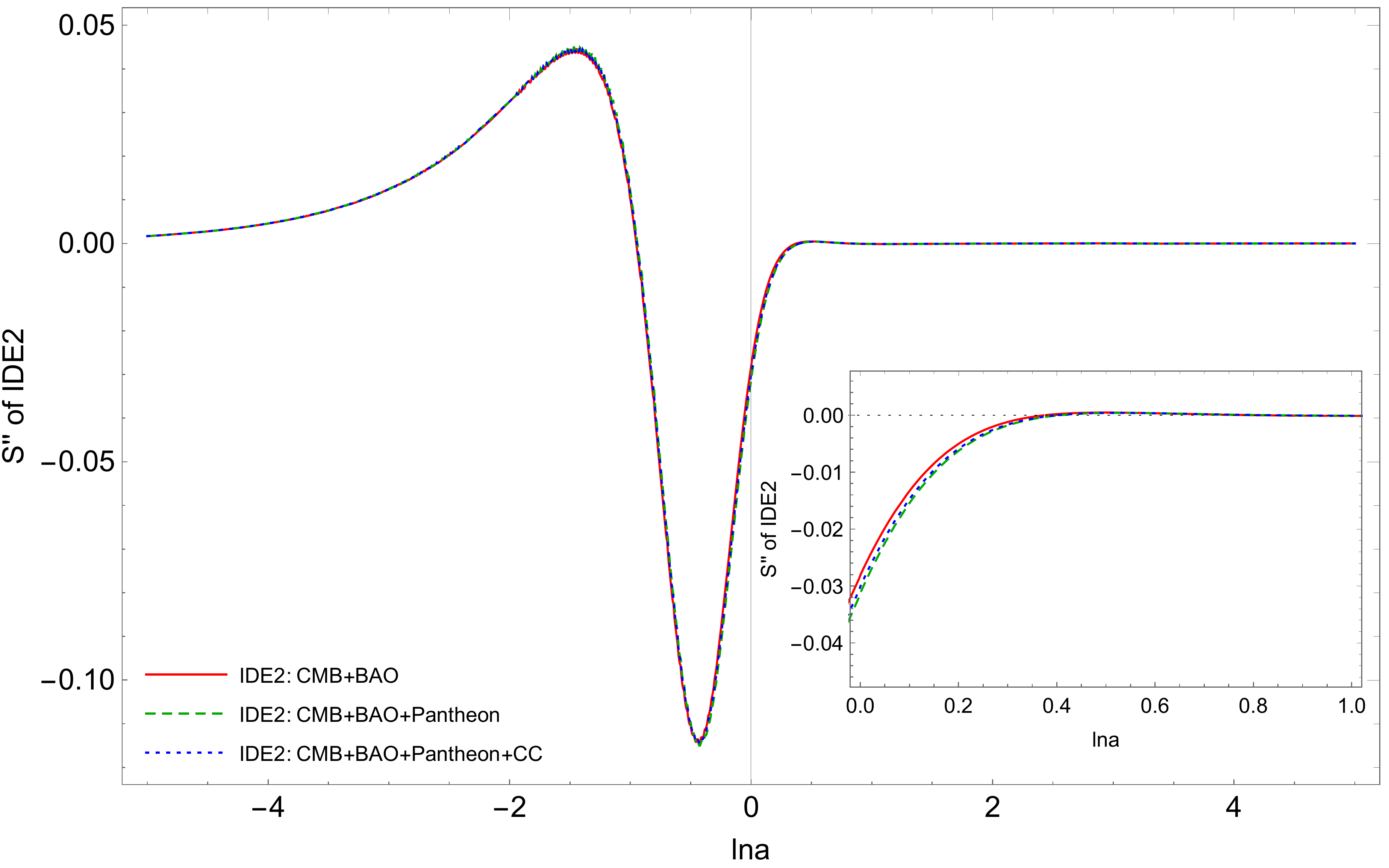}
\caption{{\it{The evolution of the second derivative of the total entropy $S^{\prime 
\prime}$ according to  (\ref{S-prime-prime}),
for IDE1 
(upper graph) and IDE2 (lower graph), for the combined analyses CMB+BAO, CMB+BAO+Pantheon 
and CMB+BAO+Pantheon+CC. In the subgraph we focus on late times, namely at  present epoch 
($\ln a = 0$) and in the future, where we can see that $S^{\prime \prime} < 0$ and thus 
the entropy is convex as required for thermodynamic equilibrium.}} }
\label{fig:thermo1}
\end{figure}

Equation 
(\ref{gslt}) can   be rewritten as 
\begin{eqnarray}\label{gslt1.1}
S^\prime = \frac{16 \pi ^2}{H^4}\, \left( H^\prime \right)^2,
\end{eqnarray}
and thus it leads to 
\begin{align}\label{S-prime-prime}
S^{\prime \prime} = 2\,S^{\prime}\, \left( \frac{ H^{\prime \prime}}{H^{\prime}}- 
\frac{2\,H^{\prime}}{H} \right) = 2\,S^{\prime}\, \left( \frac{ h^{\prime 
\prime}}{h^{\prime}}- \frac{2\,h^{\prime}}{h} \right)= 2\,S^{\prime} \Delta, 
\end{align}
where $h= H/H_0$ and $\Delta= \left( \frac{ h^{\prime \prime}}{h^{\prime}}- 
\frac{2\,h^{\prime}}{h} 
\right)$. 
In 
Fig. \ref{fig:thermo1} we depict the evolution of $S^{\prime 
\prime}$  for both  IDE1 
  and IDE2 models. As we observe, in both models at late times, namely at  present epoch 
($\ln a = 0$) and in the future,  the total entropy 
is convex, i.e.  $S^{\prime \prime} < 0$, and hence the universe tends towards 
thermodynamic equilibrium. 
This can also be seen by the fact that asymptotically $S^{\prime \prime}$ 
goes to zero. These features hold also for the two models IVS1 and IVS2.  Moreover, note 
that the individual curves  are practically 
indistinguishable.

Hence, in summary, we conclude that for cosmological scenarios with sign 
changeable interaction functions the thermodynamical laws remain 
valid.

\section{Concluding remarks}
\label{summary}

Interacting cosmology has attracted the interest of the literature, since on   one 
hand it cannot be excluded from the field theoretical point of view, and on the other 
hand it may offer a solution to the coincidence problem. However, almost all 
phenomenologically introduced interacting functions have constant sign, namely the energy 
flow maintains its direction throughout the whole universe evolution.

In the present work we investigated sign-changeable interacting scenarios, in 
which the 
interaction function, and thus the energy flow, changes sign during the evolution of the 
universe, since there is not any theoretical reason of not considering such forms. 
We  considered various   models and we extracted the involved 
equations at both the background and perturbation levels. Then we used various data 
combinations  from cosmic microwave background (CMB), 
baryon acoustic oscillations (BAO),  Supernovae Type Ia (SNIa) and cosmic chronometers 
(CC) in order to constrain the model parameters. Finally, we performed a Bayesian 
analysis in order to compare the fitting efficiency of the examined models with the 
reference  $\Lambda$CDM paradigm.

For both examined sign-changed interacting  models, namely $Q = 3 H \xi (\rho_c - 
\rho_x)$ and  $Q = 3 H (\alpha \rho_c - \beta \rho_x)$, we found that the dark-energy 
equation-of-state parameter $w_x$  prefers the phantom regime for all   datasets, at 
more than 2$\sigma$. Concerning the coupling parameters we saw that although the best-fit 
 values might be non-zero, the zero value, namely no interaction, is included within 
1$\sigma$. Moreover, we showed that this results is maintained if we impose $w_x$ to take 
 the cosmological constant value $-1$, namely considering the interacting vacuum model. 

 We proceeded by examining the effect of the interaction on the CMB TT and matter 
power spectra. As we showed, while from the simple CMB TT spectra  it is hard to 
distinguish the interacting case from $\Lambda$CDM  scenario, from the residual plot 
(with reference to $\Lambda$CDM model) one can indeed trace a distinction between the 
interacting and non-interacting cosmologies, mainly  in the lower multipoles. Similarly, 
the simple matter power spectra cannot be used to examine the interaction, however 
using the corresponding residual graphs we showed that the deviation from the 
non-interacting $\Lambda$CDM cosmology is clear.  The fact that the residual spectra 
plots can be used to distinguish the models from  $\Lambda$CDM paradigm, even if at the 
background level the latter is allowed withing 1$\sigma$ CL, is one of the main results 
of the present work.

Concerning  $H_0$, we saw that in all cases its obtained values are slightly 
higher compared to the $\Lambda$CDM-based Planck estimation, and thus the $H_0$ 
tension seems to be alleviated  as a result of the interaction, although only 
partially. Hence, we deduce that 
although the interaction may be small, it is adequate to alleviate the $H_0$ tension. 
Nevertheless, concerning   $\sigma_8$ tension   we found that the present sign 
changeable interaction models are not able to 
release it. These features show that sign-changeable interacting scenarios might be 
worthy for  further investigations.

Finally, we examined the validity of the laws of thermodynamics in a universe with 
sign-changeable interactions in the dark sector. As we saw, the generalized second law is 
always satisfied, namely the total entropy, constituting from all fluids as well as from 
the horizon entropy, is always increasing. Additionally, its second derivative becomes 
negative at late times, which implies that the universe tends towards thermodynamic equilibrium.

Last but not least, it is interesting to study the sign changeable 
models in extended parameter spaces, allowing the spatial  curvature to be non-zero. The 
inclusion of massive neutrinos could also be appealing. These projects are left for 
future investigation.

\begin{acknowledgments}
 The authors would like to thank an anonymous referee for important comments 
that 
helped   to improve the manuscript 
substantially. SP acknowledges the research funding through the Faculty Research 
and 
Professional 
Development Fund 
(FRPDF) Scheme of Presidency University, Kolkata, India. 
WY acknowledges the support from the National Natural Science Foundation of China under 
Grants No. 
11705079 and No. 11647153.
This article is based upon work from CANTATA COST (European Cooperation in Science and 
Technology) 
action CA15117, 
EU Framework Programme Horizon 2020.
\end{acknowledgments}


\end{document}